\documentclass[%
 reprint,superscriptaddress,
 amsmath,amssymb,
 aps,prx,longbibliography,]{revtex4-1}

\usepackage{graphicx}
\usepackage{dcolumn}
\usepackage{bm}

\usepackage{graphicx}
\usepackage{dcolumn}
\usepackage{bm}

\usepackage{color}

\definecolor{antiquebrass}{rgb}{0.8, 0.58, 0.46}
\definecolor{amaranth}{rgb}{0.9, 0.17, 0.31}
\definecolor{airforceblue}{rgb}{0.36, 0.54, 0.66}
\definecolor{amethyst}{rgb}{0.6, 0.4, 0.8}
\definecolor{ballblue}{rgb}{0.13, 0.67, 0.8}
\definecolor{blush}{rgb}{0.87, 0.36, 0.51}

\newcommand{\qperp}{q}
\newcommand{\bqperp}{{\bf{q}}}

\begin{document}

\title{Hyperuniform vortex patterns at the surface of type-II
superconductors}

\author{Gonzalo Rumi}
\author{Jazm\'{i}n Arag\'{o}n S\'{a}nchez}%
\author{Federico El\'{i}as}%
\author{Ra\'ul Cort\'es Maldonado}%
\author{Joaqu\'{i}n Puig}%
\author{N\'estor Ren\'e Cejas Bolecek}%
\author{Gladys Nieva}%
\affiliation{Centro At\'{o}mico Bariloche and Instituto Balseiro,
CNEA, CONICET and Universidad Nacional de Cuyo, Bariloche,
Argentina}

\author{Marcin Konczykowski}%
\affiliation{Laboratoire des Solides Irradi\'{e}s, CEA/DRF/IRAMIS,
Ecole Polytechnique, CNRS, Institut Polytechnique de Paris,
Palaiseau, France}


\author{Yanina Fasano}%
\author{Alejandro B. Kolton}%
\affiliation{Centro At\'{o}mico Bariloche and Instituto Balseiro,
CNEA, CONICET and Universidad Nacional de Cuyo, Bariloche,
Argentina}

\date{\today}

\begin{abstract}
A many-particle system must posses long-range interactions in order
to be hyperuniform at thermal equilibrium. Hydrodynamic arguments
and numerical simulations show, nevertheless, that a
three-dimensional elastic-line array with short-ranged repulsive
interactions, such as vortex matter in a type-II superconductor,
forms at equilibrium a class-II hyperuniform two-dimensional point
pattern for any constant-$z$ cross section. In this case, density
fluctuations vanish isotropically as $\sim \qperp^{\alpha}$
 at small wave-vectors $\qperp$, with $\alpha=1$. This prediction
includes the solid and liquid vortex phases in the ideal clean case,
and  the liquid  in presence of weak uncorrelated disorder. We also
show that the three-dimensional Bragg glass phase is marginally
hyperuniform, while the Bose glass and the liquid phase with
correlated disorder are expected to be non-hyperuniform at
equilibrium. Furthermore, we compare these predictions with
experimental results on the large-wavelength vortex density
fluctuations of magnetically decorated vortex structures nucleated
in pristine, electron-irradiated and heavy-ion irradiated
superconducting Bi$_{2}$Sr$_{2}$CaCu$_{2}$O$_{8 + \delta}$ samples
in the mixed state. For most cases we find  hyperuniform
two-dimensional point patterns at the superconductor surface with an
effective exponent $\alpha_{\text{eff}} \approx 1$. We interpret
these results in terms of a large-scale memory of the
high-temperature line-liquid phase retained in the glassy dynamics
when field-cooling the vortex structures into the solid phase. We
also discuss the crossovers expected from the dispersivity of the
elastic constants at intermediate length-scales, and the lack of
hyperuniformity in the $x\-y$ plane for lengths $\qperp^{-1}$ larger
than the sample thickness due to finite-size effects in the
$z$-direction. We argue these predictions may be observable and
propose further imaging experiments to  test them independently.
\end{abstract}

\maketitle


\section{\label{sec:intro}Introduction}

Hyperuniform point patterns, defined by a complete suppression of
density fluctuations in the large-wavelength
limit~\cite{TorquatoStillinger2003}, have attracted a great interest
in the last years. Such behavior can spontaneously emerge, following
either equilibrium or non-equilibrium protocols, in disordered
ground states, glass formation, jamming, Coulomb systems, spin
systems, photonic and electronic band structure, localization of
waves and excitations, self-organization, fluid dynamics, number
theory, stochastic point processes, integral and stochastic
geometry, photoreceptor cells, and even the immune
system~\cite{TorquatoStillinger2003,Torquato2018,Florescu2009}.
Hyperuniform systems are proposed to be distinguishable states of
matter characterized by special properties~\cite{Torquato2018}.
Besides, these properties can be technologically exploited directly,
or indirectly, by coupling a given system with an hyperuniform
pattern. The fabrication of such patterns, either in a controlled or
a self-assembled way, is hence also of interest from an applied
point of view~\cite{Florescu2009}.

Point patterns formed by vortex matter nucleated in type-II
superconductors have been a paradigmatic soft condensed matter
system for studying basic questions on statistical physics, such as
the statics and dynamics of elastic manifolds in random media and
glassy phases in general. This system has also been a playground to
understand the rich interplay between elasticity, quenched disorder,
thermal fluctuations, driving forces, anisotropic and finite-size
effects, either at equilibrium or out of it~\cite{LeDoussal2010}.
Since the  mean vortex density can be easily controlled by changing
the applied field $H$, vortex matter systems are particularly
suitable for studying ordering and density fluctuations at
microscopic scales, in different equilibrium or non-equilibrium
liquid, solid, or glassy phases. However, the occurrence of
hyperuniformity in vortex matter has not been experimentally
addressed yet.

Recent theoretical studies on the effect of hyperuniform pinning
arrays~\cite{LeThien2017,Sadovskyy2018} on vortex matter report an
isotropic enhancement of the critical currents in comparison with a
non-hyperuniform distribution of point pins. The work of
Ref.\,cite{LeThien2017} also predicts that vortex matter may exhibit
disordered hyperuniformity in the presence of hyperuniform or random
pinning arrays. In particular, hyperuniform vortex configurations
are proposed for rigid vortices repelling with short-range
interactions in presence of a Poisson distribution of point pins, in
a narrow region between the Bragg glass  and the vortex  liquid
phases. This last result is in contrast with the behavior expected
for the same rigid vortex system at thermal equilibrium in the
absence of disorder. Indeed, the fluctuation-compressibility theorem
forbids hyperuniformity~\cite{Torquato2018} in a system with a
non-divergent compression modulus for nearly-uniform deformations.
On the other hand,   hyperuniformity is
 expected for clean thin-film
superconductors with the penetration length much larger than the
sample thickness and logarithmic vortex-vortex repulsion.
Ref.~\cite{LeThien2017} shows that this expectation is satisfied
also in the presence of Poisson-distributed point pins.

Interestingly, as we show in this work, {\it three-dimensional}
systems made of many {\it non-rigid} interacting lines directed
along the $z$-direction are generically expected to follow an
hyperuniform point pattern for any constant-$z$ cross section of the
embedding space. This behavior arises mainly from the constraint
that elastic lines can not start nor terminate inside the medium,
and can be understood considering general hydrodynamic arguments
valid in the solid and liquid phases, even including weak pinning in
the liquid~\cite{NelsonLeDoussal1990,MarchettiNelson1991}. These
arguments show that, even though interactions between lines are
short-ranged {\it in all directions}, the effective two-dimensional
bulk modulus of the (compressible) three-dimensional system smoothly
diverges in the large-wavelength limit. Hence, at thermal
equilibrium, density fluctuations smoothly vanish and the
configurations are hyperuniform for any particular constant-$z$
slice. We show that by adding heuristic arguments these assertions
can be extended to predict the density fluctuations in the Bragg
glass, Bose glass, and liquid vortex phases with correlated disorder
generated by columnar defects (CD). We find that disorder modifies
in a non-trivial way the large-wavelength density fluctuations of
the ideal clean system. In the former case hyperuniformity becomes
marginal, while in the correlated disorder case hyperuniformity is
destroyed in the glassy and liquid equilibrium phases.
 These hydrodynamic predictions are  different from the strong
type of hyperuniformity numerically predicted in
Ref.~\onlinecite{LeThien2017} for  logarithmic and short-range
interactions in a three-dimensional system of rigid vortices in
presence of quenched disorder.

Motivated by the above predictions and open questions we
experimentally study large wavelength vortex density fluctuations in
magnetically decorated vortex structures over extended
fields-of-view (thousands of vortices) in pristine,
electron-irradiated and heavy-ion irradiated (namely with CD)
Bi$_{2}$Sr$_{2}$CaCu$_{2}$O$_{8 + \delta}$ superconducting samples.
In the observable spatial range we systematically find, for all
samples and vortex densities probed, an effective hyperuniform
behavior close to the one predicted for the line liquid in
equilibrium under weak disorder. We argue that this result can be
explained considering that larger wavelength density fluctuations
have also a slower dynamics, and are thus effectively arrested by a
realistic field-cooling process with finite temperature sweep-rate.
We also show that dispersivity of the elastic constants is
experimentally relevant, as well as finite-size effects in the
direction of the applied field, which should ultimately kill the
predicted  asymptotic hyperuniformity at in-plane scales of the
order of the superconductor thickness.

The paper is organized as follows. In Section
\ref{sec:hyperuniformity} we define the main observables that we
will use to study hyperuniformity in magnetically decorated vortex
structures. In Section \ref{sec:hydrodynamics} we review the
hydrodynamic arguments supporting the emergence  of hyperuniformity
at any $z$-constant slice of a three-dimensional system of repelling
elastic lines for different phases. In Section
~\ref{sec:experiments} we study the large-wavelength density
fluctuations in magnetically decorated vortex structures nucleated
in Bi$_{2}$Sr$_{2}$CaCu$_{2}$O$_{8 + \delta}$ samples with different
types of disorder. In Section \ref{sec:discusion} we discuss the
theoretical interpretation of these results, and report new
predictions expected for vortex matter in ad hoc experiments that
might shed more light into this issue.

\section{\label{sec:theory} Density fluctuations: Phenomenological Theory}

In this section we describe the physical magnitudes that we study
and the predictions obtained for the different vortex phases using a
hydrodynamic approach, scaling arguments and numerical simulations
of simple models.

\subsection{\label{sec:hyperuniformity} Cross-section hyperuniformity}

\begin{figure}[!ht]
\begin{center}
\includegraphics[width=\columnwidth]{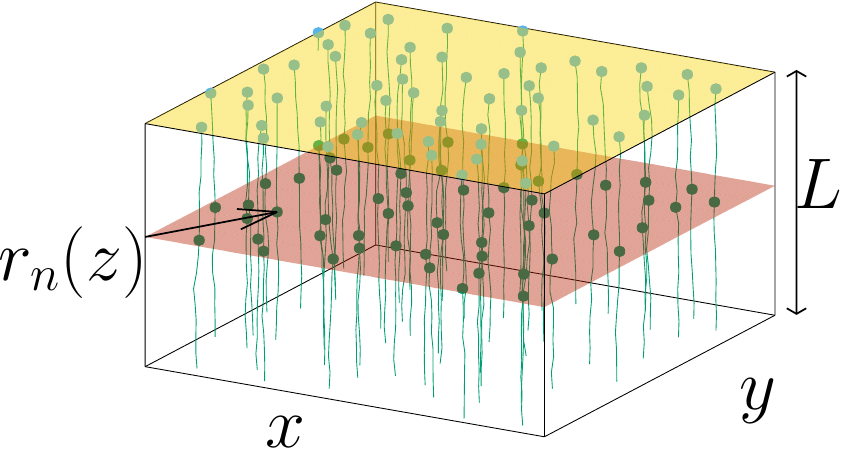}
\caption{Simulation-snapshot  of an array of fluctuating directed
elastic lines (green) modeling vortices in a type-II superconductor
with the magnetic field applied in the $z$-direction. Two
constant-$z$ cross sections are highlighted, the top and an inner
layer of the sample. The hydrodynamics of the elastic lines predicts
a class-II hyperuniform two-dimensional point pattern at each
constant-$z$ cross section (full circles), in spite of repulsion
between lines being short-ranged. The pattern frozen at the top
layer can be accessed experimentally via Bitter magnetic decoration
experiments (see Section.\ref{sec:experiments})}.
\label{fig:crosssections}
\end{center}
\end{figure}

We will be interested in the two-dimensional large-wavelength
density fluctuations at a constant-$z$ cross-section of a
three-dimensional array of vortex lines nucleated in a real sample
of thickness $L$. The average magnetic induction at equilibrium,
${\bf B} \equiv B{\bf \hat{z}}$, determines the number of vortices
per unit area in any slice, $n_0 \equiv B/\Phi_0$, with $\Phi_0=2.07
\cdot 10^{-7}$\,G.cm$^2$ the flux quantum. We model this system of
individual vortex lines by considering directed elastic lines. We
will also assume, as an approximation, that vortex lines do not
present overhangs nor pinch-off loops, and that at a given instant
the $n$th line can be appropriately described by a parameterized
position ${\bf r}_n(z)\equiv (x_n(z),y_n(z))$ (see
Fig.\ref{fig:crosssections} for a schematic representation).

The structure factor of a given frozen two-dimensional point pattern
with $N \gg 1$ points at a constant-$z$ cross section is defined as
\begin{equation}
S({\bf q},z) = N^{-1}\left|\sum_{n=1}^{N} e^{-i{\bf q}. {\bf
r}_n(z)}\right|^2 - N\delta_{{\bf q},{\bf 0}} \label{eq:Sofq}
\end{equation}

\noindent with $\mathbf{q} = (q_{x}, q_{y})$. For thick systems in
the $z$-direction $L\to \infty$, and the bulk value of
$S(\mathbf{q},z)$ is expected to be independent of $z$, namely
$S(\mathbf{q},z)\equiv S(\mathbf{q})$~\footnote{In a realistic
finite system, the $z$-dependence must be kept in order to capture
the surface-modified vortex interaction due to the stray field
interaction of vortex tips near the top $z=L$ and bottom $z=0$
layers. This effect was shown to appear only at very small $\qperp
\ll 1/L$, currently non-accessible in the typical fields-of-view of
decoration experiments. We will thus drop the $z$ dependence,
$S({\mathbf q},z)\equiv S({\mathbf q})$, and further assume that the
behavior of the top layer is representative of that of the inner
layers ~\cite{MarchettiNelson1993}.}.

A constant-$z$ cross section pattern is considered hyperuniform if
its two-dimensional large-wavelength (smallwave-vector) density
fluctuations are suppressed
\begin{equation}
\lim_{\qperp \to 0} S(\bqperp)=0.
\end{equation}
\noindent We will be particularly interested in the common cases
where the structure factor
 vanishes isotropically as a power-law near the reciprocal space origin,
\begin{equation}
S(\bqperp) \sim \qperp^\alpha, \qquad (\qperp \to 0)
\label{eq:defhyper}
\end{equation}
\noindent with $\alpha \geq 0$ a characteristic exponent. In the
asymptotic limit $N \gg 1$, this property can be translated to the
variance
 of the number $N(R)$ of particles observed inside an hyper-spherical window
with radius $R$, $\sigma_N^2(R)=\langle N^2(R) \rangle-\langle N(R)
\rangle^2$. In the latter  $\langle \dots \rangle$ denotes average
over randomly-distributed windows and the expression is valid in the
large-$R$ limit. The variance is predicted to scale as
~\cite{TorquatoStillinger2003}
\begin{eqnarray}
\sigma^2_{_N}(R) \sim \left\{
\begin{array}{lr}
R^{d-1}, \quad \alpha >1\\
R^{d-1} \ln R, \quad \alpha = 1 \qquad (R \to \infty).\\
R^{d-\alpha}, \quad 0 < \alpha < 1.
\end{array}\right.
\label{eq:variancedef}
\end{eqnarray}
For instance, a Poisson distribution of points is not hyperuniform
since the number variance is extensive with the window volume, i.e.
$\sigma_N^2 \sim R^d$. A hyperuniform system is considered to be
class-I if $\alpha>1$, class-II if $\alpha=1$, and class-III if
$0<\alpha<1$. A large and diverse list of physical and mathematical
 systems for each class is given in Ref.\,\onlinecite{Torquato2018}.


\subsection{\label{sec:hydrodynamics} Hydrodynamic arguments}

The equilibrium hydrodynamics of three-dimensional flux-line liquids
have been thoroughly discussed in the
past~\cite{NelsonLeDoussal1990,MarchettiNelson1991}. In this section
we first briefly review the basic physics and also add some new
predictions regarding large-wavelength isotropic density
fluctuations in the Bragg glass, the Bose glass (samples with
correlated disorder)  and liquid phases. This theoretical framework
is relevant for our discussion of hyperuniformity at constant-$z$
cross sections since the two-dimensional structure factor describing
large-scale density fluctuations can be  obtained from the
three-dimensional one by simple integration over the $z$-direction
wave-vectors. This yields, at equilibrium, an effective
(single-layer) two-dimensional compression
modulus~\cite{MarchettiNelson1993}.

If the penetration depth of a superconductor  is much smaller than
the thickness of the sample, $\lambda_{ab} \ll L$, (see
Fig.\ref{fig:crosssections}) vortex lines repel each other with a
roughly exponential dependence on distance for scales of order
$\lambda_{ab}$ or larger. We neglect the contribution of long-range
surface forces between vortex tips in a finite
sample~\cite{MarchettiNelson1993}). In the absence of anisotropy in
the $x-y$ plane, and for the intermediate vortex densities we are
interested in, these interactions tend to form a triangular
Abrikosov lattice well characterized by dispersive and
axially-symmetric elastic modulii of tilt $c_{44}(\qperp,q_z)$,
compression $c_{11}(\qperp,q_z)$, and  shear $c_{66}(\qperp,q_z)$.
With increasing temperature, this solid phase melts into a liquid
phase of fluctuating lines with $c_{66}(\qperp,q_z)=0$. However, in
the liquid $c_{11}(\qperp,q_z)$ and $c_{44}(\qperp,q_z)$ remain
finite and can be reinterpreted as the elastic modulii of a
viscoelastic liquid of an axially symmetric phase of elastic lines.
The three-dimensional elastic modulii depend on material parameters
and may be complicated functions of field and temperature due to the
several microscopic length-scales that come into
play~\cite{Brandt1995}, as discussed in Sec.\ref{sec:discusion}. For
the following discussion,  it is only relevant to consider that
$c_{11}(\qperp,q_z)$ and $c_{44}(\qperp,q_z)$ remain finite and
axially-symmetric in the $\qperp \to 0,\; q_z \to 0$ limit of
quasi-uniform compressions and tilting deformations. Strikingly, we
will show that, at thermal equilibrium, every constant-$z$ cross
section becomes
 incompressible, even though the three-dimensional system is compressible.

The starting point for the elastic line hydrodynamics of line liquids
is the Landau free energy functional,
\begin{eqnarray}
F&=& \frac{1}{2 n_{0}^{2}} \int d^{2} \mathbf{r} d z \int d^{2}
\mathbf{r}^{\prime} \int d z^{\prime}
\nonumber \\
&[&c_{44}\left(\mathbf{r}-\mathbf{r}^{\prime}, z-z^{\prime}\right)
\mathbf{t}\left(\mathbf{r}, z\right) \cdot \mathbf{t}\left(\mathbf{r}^{\prime}, z^{\prime}\right) \nonumber \\
&+&c_{11}\left(\mathbf{r}-\mathbf{r}^{\prime}, z-z^{\prime}\right)
\delta n\left(\mathbf{r}, z\right) \delta
n\left(\mathbf{r}^{\prime}, z^{\prime}\right)]
\nonumber \\
&+&\int d^{2} \mathbf{r} \int d z V_{D}(\mathbf{r},z) \delta
n(\mathbf{r}, z) \label{eq:freeenergy}
\end{eqnarray}
\noindent where $\mathbf{r}=(x,y)$,
$c_{44}\left(\mathbf{r}-\mathbf{r}^{\prime}, z-z^{\prime}\right)$
and $c_{11}\left(\mathbf{r}-\mathbf{r}^{\prime},
z-z^{\prime}\right)$ are the non-local tilt and compression modulii
\footnote{We are abusing the notation by using the labels $c_{11}$
and $c_{44}$ for the elastic modulii in the solid as well as in the
liquid phases. Nevertheless, in the small wave-vector
 limit, the corresponding elastic modulii are expected to be close.~\cite{Brandt1995}}
 with Fourier transforms
$c_{11}(\qperp,q_z)$ and $c_{44}(\qperp,q_z)$,
\begin{equation}
\delta n\left(\mathbf{r}, z\right)=\sum_{j=1}^{N}
\delta\left[\mathbf{r}-\mathbf{r}_{j}(z)\right]-n_0
\end{equation}
\noindent the two-dimensional vortex density fluctuations at layer
$z$ around its mean value $n_0$ and
\begin{equation}
\mathbf{t}\left(\mathbf{r}, z\right)=\sum_{j=1}^{N} \frac{d
\mathbf{r}_{j}}{d z} \delta\left[\mathbf{r}-\mathbf{r}_{j}(z)\right]
\end{equation}
\noindent the two-dimensional tangent field density for a collection
of $N \gg 1$ vortex-lines positioned at $\mathbf{r}_{j}(z)$ at a
given constant-$z$ cross section. We assume a slab geometry with a
thickness $L$ in the $z$-direction and an area $A$ in the $x-y$
plane such that $n_0 = N/A \equiv B/\Phi_0$ (see
Fig.\ref{fig:crosssections}). The last term in $F$ describes the
coupling of the vortex density with the pinning potential
$V_{D}(\mathbf{r},z)$, which can have different correlations. In
 this paper we will be interested in the cases of uncorrelated (strictly speaking
 short-range correlated) isotropic point
 disorder and the long-range correlated disorder associated to CD.

Density fluctuations around the average can be measured in the
reciprocal space by $\delta n\left(\mathbf{q},
z\right)=n\left(\mathbf{q}, z\right)-n_{0} A \delta_{\mathbf{q},
\mathbf{0}}$ where $\bqperp$ and $q_z$ are the wave-vectors in the
in-plane directions and in the average-$z$ direction of the lines.
Fluctuations are conveniently quantified by the full
three-dimensional structure factor
\begin{equation}
n_0 S^{3d}(\bqperp,q_z) = \overline{\langle |\delta
n(\bqperp,q_z)|^2 \rangle} \label{eq:defSofq3d}
\end{equation}

\noindent where $\overline{\dots}$ and $\langle \dots \rangle$
denote averages over disorder and thermal fluctuations,
respectively. These averages can be calculated, at equilibrium, from
Eq.(\ref{eq:freeenergy}) with the constraint
\begin{equation}
\partial_{z} \delta n+\nabla_{\perp} \cdot \mathbf{t}=0
\label{eq:conservation}
\end{equation}
\noindent imposing the continuity of vortex
lines~\cite{NelsonLeDoussal1990}.

It is usually
agreed~\cite{NelsonLeDoussal1990,NelsonVinokur1993,Blatter1994} that
the elastic modulii of the liquid phase $c_{11}$ and $c_{44}$ should
be similar to those of the solid or glassy phase, an assumption we
will follow in general. We will also neglect the small
renormalization of the modulii in the presence of uncorrelated
disorder~\cite{NattermannScheidl2000}. However, for correlated
disorder generated by CD, $c_{44}$ is drastically renormalized and
diverges at the Bose glass transition due to the broken statistical
symmetries. Moreover, also $c_{11}$ is expected to get strongly
renormalized and to diverge in the so-called Mott glass phase for
the field such that the number of vortices equals that of CD
(matching field). To be aware of these important differences between
these two types of disorder,
 we will denote the elastic constants in the case of correlated disorder as
${\tilde c}_{44}$ and ${\tilde c}_{11}$.

Density fluctuations are essentially controlled by compression
modes~\cite{Blatter1994}. Therefore, although the free energy of
Eq.(\ref{eq:freeenergy}) is presented for the liquid phase
($c_{66}=0$), in the ideal clean case ($V_D=0$) this expression can
be applied to compute the density fluctuations of the solid phases
as well. In real cases with disorder, for the solid phases the
situation is more subtle  since the coupling to the pinning
potential strongly depends on the periodicity  of the vortex
structure $a_0$. This implies that the $c_{66}$, not included in the
description of Eq.(\ref{eq:freeenergy}), does play a role.
Nevertheless, within the elastic approximation, hydrodynamic density
fluctuations in the solid phase with disorder can be directly
related to the prediction made for the displacement field (see
discussion in Sec.\,\ref{sec:braggglass}).

The  density fluctuations at a constant-$z$ cross section
${S}(\bqperp)$ of Eq.\ref{eq:Sofq}  can be written as
\begin{equation}
n_{0} S\left({\qperp}\right)=\overline{\langle |\delta n\left({\bqperp}, z)|^{2}\right\rangle}
\end{equation}
\noindent assuming a bulk system with statistical invariance along
the $z$-direction. ${S}(\bqperp)$ can thus be computed from
$S^{3d}(\bqperp,q_z)$ by integration over $q_z$,
\begin{equation}
S({\qperp}) \approx \frac{2\pi}{s}\int_{0}^{2\pi/s} dq_z\; {S^{3d}}({\qperp},q_z)
\end{equation}
where $s$ sets an ultraviolet cut-off coming from the discretization
in the $z$-direction or, more physically, from the superconducting
layer spacing~\cite{MarchettiNelson1993}.

\subsubsection{\label{sec:liqandsol} Liquid and Solid phases without disorder}

We start by reviewing the ideal case without disorder
 at thermal equilibrium, $V_D=0$ in Eq.\,\ref{eq:freeenergy}.
In the absence of anisotropies in the $x-y$ plane, we expect to have
roughly the same three-dimensional structure factor for the liquid,
${S}^{3d}_{\text{liq}}(\bqperp,q_z)$, and the solid Abrikosov
lattice, ${S}^{3d}_{\text{sol}}(\bqperp,q_z)$, vortex phases at
small enough wave-vectors,

\begin{equation}
{S}^{3d}_{\text{liq}}(\bqperp,q_z) \approx
S^{3d}_{\text{sol}}(\bqperp,q_z) = \frac{n_0 k_B T q^2}{q^2
c_{11}(\qperp, q_z) +q_z^2 c_{44}(\qperp, q_z)},
\label{eq:Sofq3dliquid}
\end{equation}

\noindent In this expression obtained from Eq.(\ref{eq:freeenergy})
~\cite{NelsonLeDoussal1990,MarchettiNelson1993,Radzihovsky1993,
Blatter1994},  $k_B$ is the Boltzmann constant and $T$ the bath
temperature. We are again assuming that the elastic constants  in
the vortex liquid can be well approximated by those of the solid
phase at the hydrodynamic scales ($q \cdot a_{0}<1$).
Eq.\,(\ref{eq:Sofq3dliquid}) can be obtained by relating the
spontaneous density fluctuations measured by
${S^{3d}_{liq}}({\qperp},q_z)$ with the associated linear response
function ${n_0 \qperp^2}[\qperp^2 c_{11}(\qperp, q_z) +q_z^2
c_{44}(\qperp, q_z)]^{-1}$ via the static fluctuation-dissipation
theorem. Therefore, the most important difference between the
response of the two phases, a finite $c_{66}(\qperp, q_z)$ in the
solid phase, does not play any role here. We also note that at small
wave-vectors compared to the wave-vector associated to the Bragg
peaks, $q_{0}=2\pi/a_{0}$, the solid phase has isotropic density
fluctuations since the broken rotational and translational
symmetries of this phase are irrelevant.

Neglecting for the moment  the dispersivity along the $z$-direction
(see Sec.\,\ref{sec:discusion} for a specific discussion for our
experimental case), and approximating $c_{11}$ and $c_{44}$ by their
values at $q_z=0$, we obtain following
Ref.~\onlinecite{MarchettiNelson1993}
\begin{equation}
S_{\text{liq}}({\qperp}) \approx S_{\text{sol}}({\qperp}) =\frac{n_0 k_B T}{\sqrt{c_{44}(\qperp,0) c_{11}(\qperp,0)}} {\qperp}.
\label{eq:Sofq2dliquid}
\end{equation}
\noindent for a constant-$z$ cross section of the liquid  and solid
vortex phases in a clean sample (i.e. $V_D=0$). If we assume
constant $c_{11}(\qperp,0)$ and $c_{44}(\qperp,0)$ as $\qperp \to 0$
we find that
\begin{equation}
S_{\text{liq}}(\qperp) \approx S_{\text{sol}}(\qperp)\propto \qperp.
\end{equation}
\noindent Therefore, according to the classification of
Ref.\,~\onlinecite{Torquato2018},
 in the ideal clean case the point pattern at any constant-$z$ cross
 section
is a class-II hyperuniform system ($\alpha=1$) for the liquid and
solid Abrikosov lattice phases. Some examples of systems belonging
to this hyperuniform universality class are quasicrystals, classical
disordered ground states, zeros of the Riemann zeta function,
eigenvalues of random matrices, fermionic point processes,
superfluid helium, maximally random jammed packings, perturbed
lattices, perfect glasses and density fluctuations in the early
Universe.~\cite{Torquato2018}.

\subsubsection{\label{sec:liquidwithdisorder} Weak uncorrelated disorder: Liquid phase}

The presence of disorder introduces corrections to the prediction
 of Eq.\,~(\ref{eq:Sofq2dliquid}).
Weak bulk point disorder, characterized by the correlator
$\overline{V_D({\bf r},z)V_D({\bf r'},z')}=\Delta \delta({\bf
r'}-{\bf r})\delta(z-z')$,
 produces an additive
correction to the vortex liquid three-dimensional structure factor
(Eq.~\ref{eq:Sofq3dliquid}) such that
\begin{equation}
S^{3d}_{\text{pin-liq}}(\qperp, q_z) = n_0 \Delta \left[ \frac{n_0 \qperp^2}{\qperp^2 c_{11}(\qperp,0) + q_z^2 c_{44}(\qperp,0)} \right]^2 + S^{3d}_{\text{liq}}(\qperp, q_z)
\label{eq:S3dliquidwithpointdisorder} \end{equation}
with $\Delta$ measuring the strength of the bulk disorder and where we have again
approximated the non-negligible elastic constants of the liquid by those of the
Abrikosov lattice.
This calculation neglects the dispersivity of the elastic constants along the
 $z$-direction by taking their values for $q_z=0$.
The correction to the two-dimensional structure factor is thus obtained by integrating
over $q_z$,
\begin{equation}
S_{\text{pin-liq}}(\qperp) = \frac{n_0^3 \Delta}{\sqrt{c_{44}(\qperp,0) c_{11}^3(\qperp,0)}} {\qperp} + S_{\text{liq}}(\qperp).
\end{equation}
If we again assume that $c_{11}(\qperp,0)$ and $c_{44}(\qperp,0)$
tend to constants as $\qperp \to 0$, we get
\begin{equation}
S_{\text{pin-liq}}(\qperp) \propto \qperp, \label{eq:Spinliq}
\end{equation}
\noindent as also found for the clean case. Note however, that the
prefactor is different: Only when $n_0\Delta/c_{11}(\qperp,0)< k_B
T$, the system crossovers to the line-liquid in clean samples.

This case of
 $\alpha=1$ or class-II hyperuniformity is different to the
$\alpha=2$ class-I hyperuniformity that is found for instance for
 pancake vortices interacting logarithmically
in a thin-film superconductor or the two-dimensional one-component
plasma system. The two-dimensional density fluctuations of a vortex
structure in a constant-$z$ cross section of a thick superconductor,
described by Eq.\,(\ref{eq:Spinliq}), are instead equivalent to the
ones expected for
 an effective two-dimensional particle system interacting with the
long-range Coulomb $\sim 1/r$ repulsion. Such case is indeed known
to have a two-dimensional compression modulus diverging as  $\sim
1/q$, which complies with the fluctuation-compressibility theorem.
Hyperuniformity in the liquid phase is rather robust since arises
from the continuity of flux lines, see Eq.\,(\ref{eq:conservation}).
This constraint implies that we need to deform vortex lines along
the $z$-direction in order to compress point vortices at a given
cross section, inducing a divergent effective two-dimensional
compression modulus at equilibrium. It is thus an interesting
example of hyperuniformity emerging in the three-dimensional
density-contour levels of an extended system composed by objects
with short-range repulsive interactions.

Summarizing, for constant-$z$ cross sections the vortex liquid phase
is expected to be an $\alpha=1$ class-II hyperuniform system, either
in the absence or presence of weak uncorrelated disorder.

\subsubsection{\label{sec:braggglass} The Bragg glass phase}

As mentioned, the Abrikosov lattice at thermal equilibrium is
predicted to be a class-II hyperuniform system at  constant-$z$
cross sections. Disorder-driven corrections  to the two-dimensional
structure factor  in the solid or glassy phases are subtle, and must
be considered. For  weak uncorrelated disorder  the corrections in
the solid phases are quite different than in the liquid. This is due
to the strong relevance of periodicity in the coupling of the vortex
density with disorder~\cite{Giamarchi1995,LeDoussal2010}.
Fortunately enough, if disorder is weak and the temperature low, the
elastic theory can be applied and the vortex structure can be
described by a hydrodynamic two-component displacement field
$u_{\alpha}({\bf r})$ measuring the distance of a vortex  with
respect to the corresponding perfect lattice position, such that
$u_\alpha(R_i,z)=u_{i,\alpha}(z)$ is the displacement of the $i$-th
vortex line at the slice $z$. In the case of weak uncorrelated
disorder the Bragg glass phase with quasi long-range positional
order is predicted~\cite{Giamarchi1995,Klein2001,LeDoussal2010}. The
periodic coarse-grained vortex density describing the topologically
ordered phase can be expressed as
\begin{equation}
n({\bf r},z)= n_0 \left( 1 - \partial_{{\bf r}}.{\bf u}({\bf
r},z)+...\right),
\end{equation}
\noindent where the ``$\dots$'' denote rapidly oscillating terms
that do not contribute to the hydrodynamic density modes we are
interested in. The hydrodynamic three-dimensional structure factor
is
\begin{equation}
S^{3d}_{sol}(\qperp,q_z) \approx n_0 \qperp^2 \overline{\langle
|u_L({\bf q},q_z)|^2 \rangle}. \label{eq:Sfromu}\end{equation}
\noindent where $u_L(\bqperp,q_z)$ is the component of the
displacement field in the direction parallel to $\bqperp$. This
expression is rather general, see Ref.\,\onlinecite{Torquato2018b}.
If we consider the Abrikosov lattice corresponding to the clean
system below the melting temperature, the thermal roughening of such
displacement component is
\begin{equation}
\overline{\langle |u_L(\bqperp,q_z)|^2 \rangle}
\sim
\frac{k_B T}{\qperp^2 c_{11}(\qperp, q_z) +q_z^2 c_{44}(\qperp, q_z)},
\end{equation}
\noindent The structure factor of the lattice is thus essentially
the same as for the liquid,
\begin{equation}
S^{3d}_{\text{sol}}({\qperp},q_z) \approx  \frac{n_0 k_B T
{\qperp^2}}{\qperp^2 c_{11}(\qperp, q_z) +q_z^2 c_{44}(\qperp, q_z)}
\approx S^{3d}_{liq}(\qperp,q_z), \label{eq:Sofq3dbraggglas}
\end{equation}
\noindent since the structure factor in both phases is controlled by
the longitudinal modes only. At finite temperatures, the Abrikosov
lattice at constant-$z$ hence displays the same $\alpha=1$
hyperuniformity as the liquid,
\begin{equation}
S_{\text{sol}}(\qperp) \propto \qperp,
\end{equation}
\noindent as already discussed in the previous section.

Weak disorder destroys the perfect long-range positional order,
 but not the topological order, and a particular quasi-long range order emerges.
 In this Bragg glass phase~\cite{Giamarchi1995,LeDoussal2010} the displacement field
grows as a power-law at intermediate distances, and logarithmically
at large distances, in contrast with the temperature-dependent
saturation of displacements found in the  Abrikosov lattice
(associated to the Debye-Waller factor). The disorder-induced
roughening of the displacement field is expected to be isotropic in
the $x-y$ plane at large-distances with its corresponding structure
factor, see Eq.\,\ref{eq:Sfromu}, to reduce to the clean thermal
case of Eq.(\ref{eq:Sofq3dliquid}) on vanishing disorder.
Heuristically, we can thus approximate the thermal- and
disorder-induced displacements as
\begin{equation}
\overline{\langle |u_L(\qperp,q_z)|^2 \rangle}
\sim
(\qperp^2 {c_{11}}(\qperp, q_z) +q_z^2 {c_{44}}(\qperp, q_z))^{-\frac{d+2\zeta}{2}},
\end{equation}
\noindent with $d=3$ the space dimension and $\zeta$ a
characteristic exponent. Note that we have used the same ${c_{44}}$
and ${c_{11}}$ as for the clean case, assuming no disorder-induced
renormalization. In presence of thermal noise we have $\zeta
=1-d/2$, thus reducing to Eq.(\ref{eq:Sofq3dliquid}). When $\zeta>0$
this exponent is the so-called roughness exponent of the lattice,
such that the mean square displacement grows as
\begin{equation}
\langle u_L^2 \rangle \sim \int d\qperp\; dq_z\;\qperp^{d-2}
\overline{\langle |u_L(\bqperp,q_z)|^2 \rangle} \sim R^{2\zeta}
\end{equation}
\noindent with $R$ the observation scale or system linear size in
the $x-y$ plane. For $\zeta=0$ the lattice is logarithmically rough
$\langle u_L^2 \rangle \sim \log R$, while for $\zeta<0$ has a
macroscopically flat displacement field.

By integrating  over $q_z$ and assuming consistently that the
elastic constants ${c_{11}}$ and ${c_{44}}$ tend to finite values at
small $q_z$, we obtain the constant-$z$ cross section
two-dimensional structure factor of the pinned solid,
$S_{\text{pin-sol}}(\qperp) \sim \qperp^{3-d-2\zeta}[c_{44}
c_{11}^{d+2\zeta-1}]^{-1/2}$ and thus, for $d=3$
\begin{eqnarray}
S_{\text{pin-sol}}(\qperp) \sim \frac{\qperp^{-2\zeta}}{\sqrt{
c_{44}(\qperp,0)  c_{11}^{2+2\zeta}(\qperp,0)}}
\label{eq:Sofq2dbraggglas}
\end{eqnarray}
\noindent Therefore, under these assumptions, the elastic system is
hyperuniform only if $\zeta<0$ (displacement field macroscopically
flat) while for $\zeta=0$ is marginally hyperuniform, i.e.
$\alpha=0$. This is particularly important for the Bragg glass and
for systems with quasi-long-range order (such as two-dimensional
crystals with short-range interactions), since $\zeta \to 0$
asymptotically. This so-called random-periodic regime (RP) of the
Bragg glass phase exists for $\qperp R_a \ll 1$, where $R_a$ is the
scale at which the displacement field has fluctuations of order
$a_0$. For $R_c<R<R_a$, where $R_c$ is the Larkin radius on the
$x-y$ plane, there is a crossover to the so-called random-manifold
regime (RM) where $\zeta \equiv \zeta_{RM} \approx 0.2$. For scales
shorter than  $R_c$, the system crossovers to the Larkin regime,
where $\zeta \equiv \zeta_L = (4-d)/2 = 1/2$. In none of these
regimes the Bragg glass phase presents suppressed density
fluctuations.

Summarizing, any arbitrarily weak disorder kills the class-II
hyperuniformity of the Abrikosov lattice, and the Bragg glass phase
is expected to be marginally hyperuniform at constant-$z$ cross
sections. Rather surprisingly, from the given arguments, the Bragg
glass is expected to jump from marginal to a class-II
hyperuniformity at melting, adding a new signature to this
first-order phase transition~\cite{Giamarchi1995,LeDoussal2010}.

\subsubsection{\label{sec:columnardisorder} CD correlated disorder: Liquid and Bose glass}

The case of the correlated disorder generated by
randomly-distributed CD, such that $\overline{V_D({\bf r},z)
V_D({\bf r'},z')}=\Delta_1 \delta({\bf r'}-{\bf r})$, is special
since the long-range correlation along $z$ favors the localization
of vortex lines into CD. A Bose glass
 transition is expected by lowering the temperature from the liquid
 phase~\cite{NelsonVinokur1993}, with the
  concomitant divergence of ${\tilde c}_{44}$ approaching the transition temperature,
  while ${\tilde c}_{11}$ remains finite. There is an exception for
  the putative Mott glass phase
  at the matching field $B=B_\Phi \equiv n_{\text{col}}\Phi_0$
where ${\tilde c}_{11}$ is also expected to diverge. The predicted
three-dimensional structure factor for the vortex liquid in presence
of   CD disorder is
\begin{eqnarray}
S^{3d}_{\text{col-liq}}(\bqperp,q_z) &\approx& \frac{n_0 k_B T \qperp^2}{\qperp^2 {\tilde c}_{11}(\qperp, q_z) +q_z^2 {\tilde c_{44}}(\qperp, q_z)},
\nonumber \\
&+&\Delta_1 \frac{n_0^3}{{\tilde c_{11}^2}(\qperp,0)}\delta(q_z)
\label{eq:Sofq3dboseliquid}
\end{eqnarray}
\noindent where ${\tilde c_{11}}$ and ${\tilde c_{44}}$ are the
compression and tilt modulii in the presence of correlated CD
disorder, and $\Delta_1$ measures the strength of the columnar
disorder. Integrating over $q_z$ we obtain the two dimensional
structure factor
\begin{eqnarray}
{S}_{\text{col-liq}}(\qperp) &\approx&
\frac{n_0 k_B T}{\sqrt{{\tilde c_{44}}(\qperp,0) {\tilde c_{11}}(\qperp,0)}} \qperp
\nonumber \\
&+& \Delta_1 \frac{n_0^3}{{\tilde c_{11}^2}(\qperp,0)}.
\label{eq:Sofq2dboseliquid}
\end{eqnarray}
\noindent This expression yields a crossover wave-vector
\begin{equation}
\qperp_{CD} = \frac{n_0^2 \Delta_1}{{\tilde c_{11}} k_B T}
\sqrt{\frac{{\tilde c_{44}}}{{\tilde c_{11}}}} \label{eq:qcd}
\end{equation}
\noindent such that for scales $q<q_{CD}$, the density fluctuations
in the liquid are dominated by the pinning introduced by CD (second
term in Eq.\,\ref{eq:Sofq2dboseliquid}),
 while for $q>q_{CD}$ they are similar to those found in the liquid phase
 without correlated CD disorder (first term in Eq.\,\ref{eq:Sofq2dboseliquid}).

On approaching the Bose glass transition  on cooling,
${\tilde c_{44}}$ increases rapidly and
diverges at the transition, concomitantly with
$q_{CD} \to \infty$. We hence expect
\begin{eqnarray}
{S}_{\text{col-sol}}(\bqperp) &\approx&
\Delta_1 \frac{n_0^3}{{\tilde c_{11}^2}(\qperp,0)}.
\label{eq:Sofq2dboseglass}
\end{eqnarray}
\noindent in the Bose Glass phase where the localized vortex lines
repel each other with short-range interactions. The compression
modulus remains finite at the transition~\cite{NelsonVinokur1993}
and therefore a non-hyperuniform system is expected at equilibrium.

Summarizing, correlated disorder generated by CD destroys
hyperuniformity at constant-$z$ cross sections, both in the Bose
glass and in the liquid phases. However, since on increasing
temperature $q_{CD}$ vanishes, a crossover to a nearly class-II
hyperuniform liquid can occur within the liquid phase. This
crossover can also occur on decreasing the magnitude and/or density
of correlated disorder $\Delta_{1}$.

\subsubsection{\label{sec:finitesizeeffects} Finite-size effects}

In the previous subsections, we assumed infinite samples and showed
that the ideal clean liquid and solid phases, as well as the liquid
phase with weak disorder, are class-II hyperuniform systems. In
samples with thickness $L$, finite effects in the $z$-direction will
affect hyperuniform behavior inducing a crossover at a
characteristic $\qperp_{FS}(L)$. Indeed, by a simple dimensional
analysis of the three-dimensional structure factors in these
hyperuniform cases (Eqs.\,\ref{eq:Sofq3dliquid} and
\ref{eq:S3dliquidwithpointdisorder}), we get
\begin{equation}
\qperp_{FS} \sim
\sqrt{\frac{c_{44}(\qperp_{FS},2\pi/L)}{c_{11}(\qperp_{FS},2\pi/L)}}\frac{2\pi}{L}.
\label{eq:qstar}
\end{equation}
\noindent Physically, $\qperp_{FS}$ is the characteristic
wave-vector at which the correlation length of vortices in the
$z$-direction becomes of order $L$~\cite{MarchettiNelson1993}. In
order to illustrate this, for simplicity we perform the calculation
from Eq.\,\ref{eq:Sofq3dliquid} for the clean system. By Fourier
inverting in the
 $z$-direction,
\begin{equation}
S^{3d}(\bqperp,z_1-z_2) = \frac{n_0 k_B T}{c_{11}(\qperp,0)\xi_\parallel(\qperp)}e^{-|z_1-z_2|/\xi_{\parallel}(\qperp)}
\end{equation}
with
\begin{equation}
\xi_{\parallel}(\qperp)=\qperp^{-1}
\sqrt{c_{44}(\qperp,2\pi/L)/c_{11}(\qperp,2\pi/L)}.
\end{equation}
\noindent Therefore, Eq.\,\ref{eq:qstar} is equivalent to the
physical condition $\xi_{\parallel}(\qperp_{FS})=L$. For $\qperp <
\qperp_{FS}$ the three-dimensional system essentially behaves as a
two-dimensional system of rigid lines with the structure factor
obtained by setting $z_1=z_2$, and $\xi_\parallel = L$,
\begin{equation}
{S}(\bqperp) = \frac{n_0 k_B T}{c_{11}(\qperp,0)L}.
\label{eq:finitesizeSofq}
\end{equation}
\noindent Provided $L>\lambda_{ab}$ at the crossover, rigid vortices
repel with a short-range interaction
yielding~\cite{Blatter1994,NattermannScheidl2000}
\begin{equation}
c_{11}(q)=2 \epsilon_{0} \frac{2 \pi
\lambda_{ab}^{2}}{1+\lambda_{ab}^{2} q^{2}}
\frac{B^{2}}{\Phi_{0}^{2}}. \label{eq:c11rigidvortices}
\end{equation}
\noindent with $\epsilon_0 = (\Phi_0/4\pi \lambda_{ab})^2$ the line
tension. Therefore, since the length-scale in the $x-y$ plane probed
is $q^{-1} \gg \lambda_{ab}$, we have ${S}(\bqperp) \to
\text{const}$ as $\qperp \to 0$. In other words, finite-size effects
kill the class-II hyperuniformity of the three-dimensional vortex
liquid or solid phases in clean samples at thermal equilibrium.
Although this result was obtained for a particular case, this
finite-size effect is expected to be present in all vortex phases,
even with disorder. Indeed, the existence of $\qperp_{FS}(L)$ comes
from first, a dimensional analysis of the competition between
tilting and compression elastic responses, and second, from the
non-dispersive behavior of the three-dimensional elastic constants.

\subsubsection{\label{sec:twodimensional} Two-dimensional systems}

It is interesting to discuss Eq.\,\ref{eq:finitesizeSofq} in the
limit of very thin superconductors such that $L \ll \lambda_{ab}$.
In this limit the effective penetration length becomes $\Lambda =
2\lambda_{ab}^2/L$. If the observation length-scale in the $x-y$
plane is $q \Lambda > 1$, two different power-law behaviors can be
expected. At very small length-scales,  $q \Lambda \gg 1$,
interactions are logarithmic and
\begin{equation}
c_{11}(\mathbf{q})=2 \epsilon_{0} \frac{2 \pi}{q^{2}}
\frac{B^{2}}{\Phi_{0}^{2}}
\end{equation}
\noindent in Eq.\,\ref{eq:finitesizeSofq}, so the system at
equilibrium is class-I hyperuniform with $\alpha=2$,
\begin{equation}
{S}(\qperp) \to \qperp^2
\end{equation}

\noindent for $\log(1/x)$ interactions. Recent simulations with this
type of  interaction show that the value $\alpha=2$ is robust under
the presence of quenched uncorrelated disorder.~\cite{LeThien2017}

For larger length-scales, $q \Lambda > 1$, the interaction becomes
Coulomb-like  and
\begin{equation}
c_{11}(\mathbf{q})=2 \epsilon_{0} \Lambda \frac{2
\pi}{q}\left(\frac{B}{\Phi_{0}}\right)^{2},
\end{equation}
\noindent yielding, at equilibrium, a class-II hyperuniform vortex
pattern with $\alpha=1$  and thus
\begin{equation}
{S}(\bqperp) \to \qperp
\end{equation}
\noindent for $1/x$ interactions. Interestingly, this is the
constant-$z$ cross section behavior predicted for the
three-dimensional line array. In other words, the effective
two-dimensional interaction at a constant-$z$ cross section,
mediated by the short-range three-dimensional interaction between
vortex lines, mimics Coulomb interactions (as in Wigner crystals).
This is analogous to the long-range elasticity of the triple contact
line of a liquid meniscus,
 so called ``fringe elasticity of the line of contact'', reflecting the
 energetics of deformations of the liquid/gas
 interface~\cite{Joanny1984,degennes2003}.
Although we are assuming clean systems, since   for the
two-dimensional elastic lattice in the presence of weak disorder
$c_{11}$  is not
renormalized~\cite{Carpentier1997,NattermannScheidl2000},
 we can expect the
result $\alpha=1$ to hold even in
the presence of weak uncorrelated pinning. This would be the case
 in the solid as well as in the liquid two-dimensional vortex structure,
due to their similar uniaxial compression properties at small
wave-vectors.

\subsubsection{\label{sec:summarypredictions}
Summary of analytical predictions}

In Table \ref{tab:predictions} we summarize our predictions
regarding the type of hyperuniformity of different vortex phases at
constant-$z$ cross sections of a three-dimensional superconductor
(disregarding the finite-size effects discussed above). Liquid
vortex phases are expected to be class-II hyperuniform for weak
uncorrelated pinning, while correlated disorder destroys
hyperuniformity. As discussed, a crossover of the liquid with CD
correlated disorder towards an apparent class-II hyperuniformity at
large temperatures or weak disorder is possible. At low
temperatures, only the Abrikosov lattice with thermal fluctuations
is class-II hyperuniform. The Bragg glass phase, expected for weak
uncorrelated disorder, is marginally hyperuniform due to the
disorder-induced roughening of the displacements field, while the
Bose glass phase is not hyperuniform due to the enhanced
correlations along the $z$-axis. The putative Mott glass phase is
expected to be hyperuniform due to the divergence of both, the tilt
and the compression modulii. To the best of our knowledge, the way
${\tilde c}_{11}$ diverges with $\qperp$ has not been reported for
the Mott glass, and hence we can not guess its hyperuniformity class
yet. All these predictions assume three-dimensional vortex line
phases, non-dispersive elastic constants in the low $\qperp$ limit,
and neglect finite-size effects.

\begin{table}[h]
  \centering
\begin{tabular}{|c|c|}
\hline
{\it Phases} & {\it hyperuniformity } \\
\hline
Liquid without disorder & {\bf Yes} : Class II $\alpha=1$ \\
\hline
Liquid with weak uncorrelated disorder & {\bf Yes} : Class II $\alpha=1$ \\
\hline
Liquid with correlated disorder by CDs & {\bf No} \\
\hline
Abrikosov crystal & {\bf Yes}: Class II $\alpha=1$  \\
\hline
Bragg Glass & {\bf Marginal}: $\alpha=0$\\
\hline
Bose Glass & {\bf No} \\
\hline
Mott Glass & {\bf Yes}: Unknown Class. \\
\hline
\end{tabular}
\caption{Analytical predictions for the type of hyperuniformity for
different three-dimensional vortex-line phases at constant-$z$ cross
sections in the thermodynamic limit. } \label{tab:predictions}
\end{table}

\subsection{\label{sec:simulations}Numerical Simulations}

The hydrodynamic predictions presented in the previous sections are
based on a coarse-grained continuous model and the assumption of
thermal equilibrium. In order to test these analytical predictions,
and to match two different levels of description of the problem, we
have performed molecular dynamic simulations of a simple microscopic
model. This helps us to understand finite-size and discretization
effects in the $z$-direction, as well as the effect of the
particular shape of the vortex-vortex interaction potential, without
relying in the approximated elastic modulii of the continuum
description.

We model vortices as elastic lines discretized in the $z$-direction,
such that ${\bf r}_{i}(z) \equiv (x_i(z),y_i(z))$ describe their
two-dimensional coordinates at the layer $z$, with $z=0,...,N_z-1$
and $N_z$ the total number of layers. Periodic boundary conditions
are considered in all directions. The total energy of the elastic
lines array is $E[\{{\bf r}_{i}(z)\}]=E_1 + E_{int}$ such that each
line has an elastic tension energy given by Hook coupling with
strength $k$,
\begin{equation}
E_{1}=\sum_{i=1}^{N} \sum_{z=0}^{N_{z} - 1} \frac{k}{2} |{\bf
r}_{i}(z+1)-{\bf r}_{i}(z)|^2
\label{eq:tensionv}
\end{equation}
\noindent and the repulsive interaction energy between vortex-lines
is modeled as
\begin{equation}
E_{int}=\sum_{i \neq j} \sum_{z=0}^{N_{z} - 1} {\epsilon_0} K_0
\left( \frac{|{\bf r}_{j}(z)-{\bf r}_{i}(z)|}{\lambda_{ab}} \right)
\label{eq:intvv}
\end{equation}
\noindent with $K_0(x)$ the zero-order modified Bessel function of
the second kind. Eq.\,\ref{eq:intvv} is derived from the London
model, while Eq.\,\ref{eq:tensionv} with $k \propto \epsilon_0$ is a
harmonic local approximation for the single-vortex elastic tension.
In layered superconductors as the one experimentally study here, the
elastic tension energy arises from the attractive electromagnetic
and Josephson couplings between pancakes of different
superconducting layers. We consider an over-damped Langevin dynamics
at a temperature $T$
\begin{eqnarray}
&\eta& \partial_t {\bf r}_{i}(z) = -\frac{\delta E}{\delta {\bf r}_{i}(z)} + \xi({\bf r}_{i}(z),t) \\
&\langle& \xi({\bf r}_{i},t) \xi({\bf r'}_{j},t') \rangle = 2 \eta
k_B T \delta_{ij}  \delta(t-t')
\end{eqnarray}

\noindent where $\eta$ is the Bardeen-Stephen friction. At long
enough times this system equilibrates in the canonical ensemble at
temperature $T$. In order to approach experimental conditions we
have simulated a range of $a_0$ close to the typically accessed in
magnetic decorations.

\begin{figure}[t]
 \includegraphics[width=0.84\columnwidth]{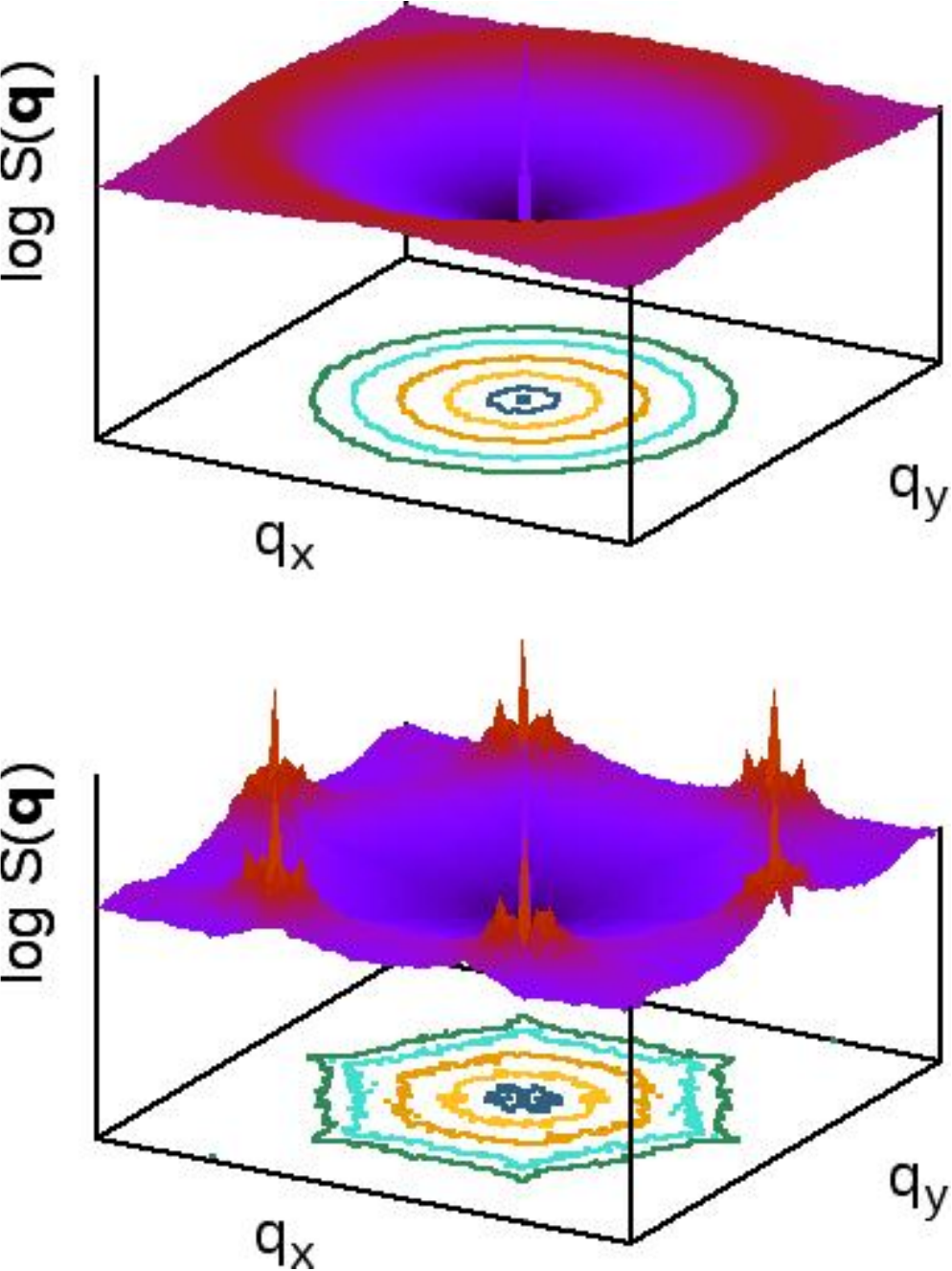}
 \caption{ Typical structure factors
 for a liquid and a solid at equilibrium at finite $T$ for an ideal
 clean superconductor, as obtained from
 Molecular Dynamics simulations of an elastic-line array with short-ranged elasticity
 and repulsive short-ranged interactions. Axial symmetry
 in both phases is observed in the large wavelength (low $q$) limit.}
 \label{fig:lowqiso}
\end{figure}

\begin{figure}[ttt]
 \includegraphics[width=0.9\columnwidth]{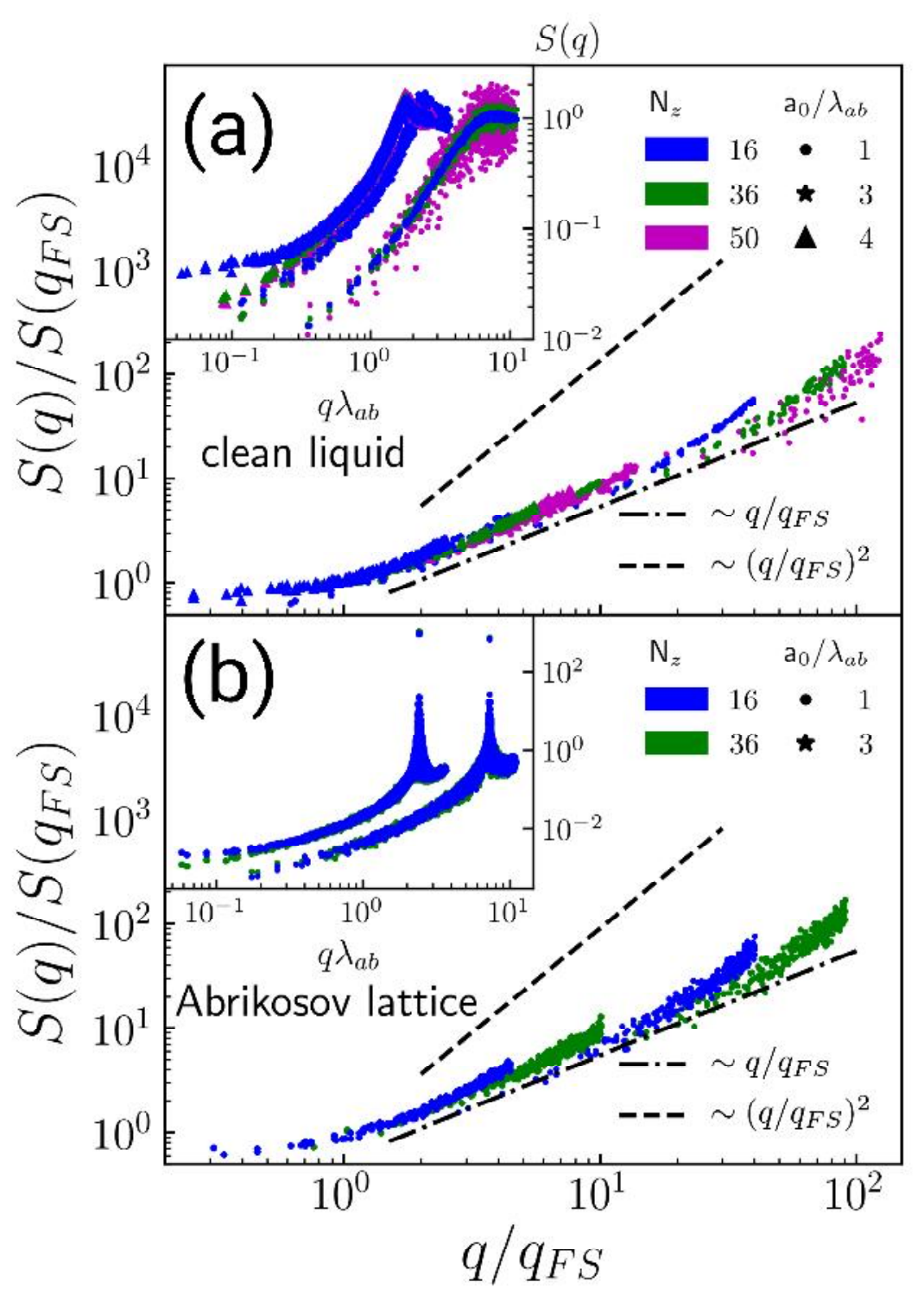}
 \caption{Normalized two-dimensional structure factor obtained from
 Molecular Dynamics simulations of a three-dimensional elastic-line array
 in the ideal clean case, for various lattice spacings $a_0$ and number
 of layers $N_z$. We use a finite-size crossover scale
  $q_{FS} \propto a_0/N_z$ and a normalization $S(q_{FS})\propto T
  a_0^2/N_z$.  (a) Liquid phase at equilibrium and (b) solid Abrikosov phase.
 Both phases display the same finite-size crossover at $q_{FS}$.
 For $q<q_{FS}$ the system effectively becomes a non-hyperuniform two-dimensional
 system, while for $q>q_{FS}$ it shows a fair $S(q)\sim q$ hyperuniform behaviour
 at low $q$.
 Upward corrections at large wave-vectors, $q \lambda_{ab} \gtrsim 1$, can
 be explained by the dispersivity of the compression modulus emerging from
  the particular interaction potential. Lines are a guide to the eye for $S(q)\sim q^{\alpha}$ with
 $\alpha=2$ (dashed line) and $\alpha=1$ (dashed-dotted line).
 The insert shows non-normalized data.}
 \label{fig:liquidsimulation}
\end{figure}

Figure\,\ref{fig:lowqiso} shows typical two-dimensional structure
factors $S(\mathbf{q})$ averaged over the $N_z$ layers. The liquid
and solid without disorder in thermal equilibrium, have both an
isotropic structure factor at low wave-vectors, although at large
wave-vectors the solid displays the peaks corresponding to the
triangular Abrikosov lattice. This isotropy at low $q$ is consistent
with the hydrodynamic assumption made in
Sec.\,\ref{sec:hydrodynamics} considering axially-symmetric elastic
constants. Thus, for the $q$-range we work with the angular average
of the structure factor, $S(q)\equiv (2\pi)^{-1}\int_0^{2\pi} d\Psi
S(q\cos(\Psi), q\sin(\Psi))$, in order to examine large length-scale
density fluctuations.

The inserts of Fig.\,\ref{fig:liquidsimulation}   show $S(q)$ for a
line liquid above the melting temperature and for an ordered line
solid, both at equilibrium and without disorder. The figure shows
results for different lattice spacings $a_0/\lambda_{ab}=1,3,4$,
where $A=N a_0^2 \sqrt{3}/2$ is the area of the computational box,
and different number of layers $N_z=16,36,50$ in the $z$-direction.
There are clear finite size effects, i.e. $N_z$ dependence, in the
$z$-direction, and they are amplified when $a_0$ increases. We find
that these size effects can be quantified by appropriately
normalizing the inset curves into a master curve: we consider a
characteristic finite-size crossover scale $q_{FS}$ such that
$S(\qperp)/S(q_{FS})\sim  G(\qperp/q_{FS})$, with $G(x)\sim 1$ for
$x\ll 1$ and a unique $G(x)$ for $x \gg 1$. For the range of $a_0$
analyzed, we find a good collapse in the low-$q$ region using the
crossover wave-vector $q_{FS}=a_0/N_z$, and $S(q_{FS}) \propto T
a_0^2/L$, both for the liquid and solid phases. Therefore, the
saturation of $S(q)$ for $q<q_{FS}$ is actually a finite-size
effect, and it should disappear in the thermodynamic $N_z \to
\infty$ limit. The dependence of $q_{FS}$ with $N_z \propto L$
confirms the size effects predicted in
Sec.\,\ref{sec:finitesizeeffects}. In addittion, the shape of the
master curve  for $q>q_{FS}$, $G(x) \sim x$, confirms an $\alpha
\approx 1$ typical of class-II hyperuniformity (see
Eq\,\ref{eq:Sofq2dliquid}).

The evolution of $q_{FS}$ with $a_0$ is less universal than the
dependence with $N_z$ since relies on the precise functionality of
the elastic modulii $c_{44}$ and $c_{11}$ that depend on the type of
inter-vortex interaction and the field regime explored in the
simulations. However, if we use the rigid vortex compression modulus
of Eq.\,\ref{eq:c11rigidvortices}, $c_{11}(\qperp,q_z=0) \propto
B^2/(1+\qperp^2 \lambda_{ab}^2)$, the single-vortex tilt modulus
$c_{44} \propto \epsilon_0 \sim B^2 a_0^2$ (a constant in this
particular model), and approximate Eq.\,(\ref{eq:qstar}) as $q_{FS}
\approx \sqrt{c_{44}/c_{11}(0,0)}/L$, we get $q_{FS} \propto a_0/L$.
Also, from Eq.\,\ref{eq:finitesizeSofq} we get $S(q_{FS}) \sim n_0
k_B T/c_{11} L \propto T a_0^2/N_z$. Both results are in agreement
with the outcome of simulations shown in
Fig.\ref{fig:liquidsimulation} confirming that the $c_{11}$ for
rigid vortices  is adequate for describing the $S(q)$ saturation for
$q<q_{FS}$.

For $q > q_{FS}$ the master curve presents deviations from $S(q)
\sim q$ for  $q \gg q_{FS}$, crossing over to a more rapid increase
of $S(q)$ with $q$. This behavior can be also qualitatively
explained in terms of the hydrodynamic prediction of
Eq.\,\ref{eq:Sofq2dliquid} if we consider that $S(q) \propto
\qperp/\sqrt{c_{44}c_{11}(\qperp,0)} \sim \qperp
\sqrt{1+\lambda_{ab}^2 \qperp^2} \approx \qperp(1+\qperp^2
\lambda_{ab}^2/2+\dots)$. Therefore, this microscopic model confirms
the existence of finite-size effects, and the class-II
hyperuniformity of the liquid and solid vortex-line phases in ideal
clean samples. We also explain the upward deviations from the
$\alpha=1$ behavior observed at  $q > q_{FS}$ controlled by the
dispersive behavior of $c_{11}$, emerging when $q\lambda_{ab} \sim
1$. As discussed later, dispersive effects may be important for the
range of parameters given by the magnetic decoration experiments in
the layered superconductor Bi$_2$Sr$_2$CaCu$_2$O$_{8 + \delta}$. The
 study presented in this section bridges the hydrodynamic scale with the molecular dynamics
scale, using a simple elastic-line array model.

\section{\label{sec:experiments}Density fluctuations: Experimental Results}

In order to test the previous theoretical predictions we have
performed magnetic decorations in various superconducting samples
giving us direct access to two-dimensional point
 patterns at the top surface.
This layer is the particular $z=L$ cross section where surface
interactions may play an important role. However, it has been argued
that the structure at the surface is  representative of that at the
bulk of the sample for the typical in-plane length-scales probed by
decorations $\sim a_0$~\cite{MarchettiNelson1993}. An alternative
experimental method to the one described here would be to study
small-angle neutron scattering data, yielding directly the structure
factor. We followed this path but found that the experimental
resolution of this technique at  low $q$ is not good enough as to
ascertain whether hyperuniformity is present in the vortex system.

The choice of the extremely layered
Bi$_2$Sr$_2$CaCu$_2$O$_{8+\delta}$  system is based in the easiness
to obtain freshly cleaved surfaces to perform several decoration
experiments in the same crystal, and on the availability of samples
with different types of disorder (pinning potentials). We studied
vortex lattices nucleated on a large set of single-crystals with the
natural disorder coming from crystalline defects (pristine samples),
and with additional disorder introduced by irradiation with electron
and heavy ions. Electron-irradiated samples present a
 dense point-like disorder whereas heavy ion-irradiated ones
 present a Poisson-like distribution of CD, columns of crystallographic
 defects aligned along the $z$-direction. Electron irradiation was performed with
2.3\,MeV accelerated electrons in a van de Graaff accelerator
coupled to a closed-cycle hydrogen liquifier at the \'{E}cole
Polytechnique of Palaiseau, France~\cite{Konczykowski2009}. This
process is performed at low temperatures ($20$\,K) as to guarantee
the stability of Frenkel pairs created in the irradiation process.
The data presented here corresponds to a sample irradiated with an
electron density of
$1.7\,10^{19}$\,e/cm$^2$~\cite{Konczykowski2009}. The irradiation of
samples with energetic ($\sim 1$\,GeV) heavy ions of Xe and Pb
resulted in samples with correlated CD disorder characterized by the
matching field $B_{\Phi}$. The studied samples have a low density of
CD: $B_\Phi=30$ and $60$\,Gauss for the Xe-irradiated samples and
$B_\Phi=45$ and $100$\,Gauss for the Pb-irradiated ones.

\begin{figure}[ttt]
    \includegraphics[width=0.75\columnwidth]{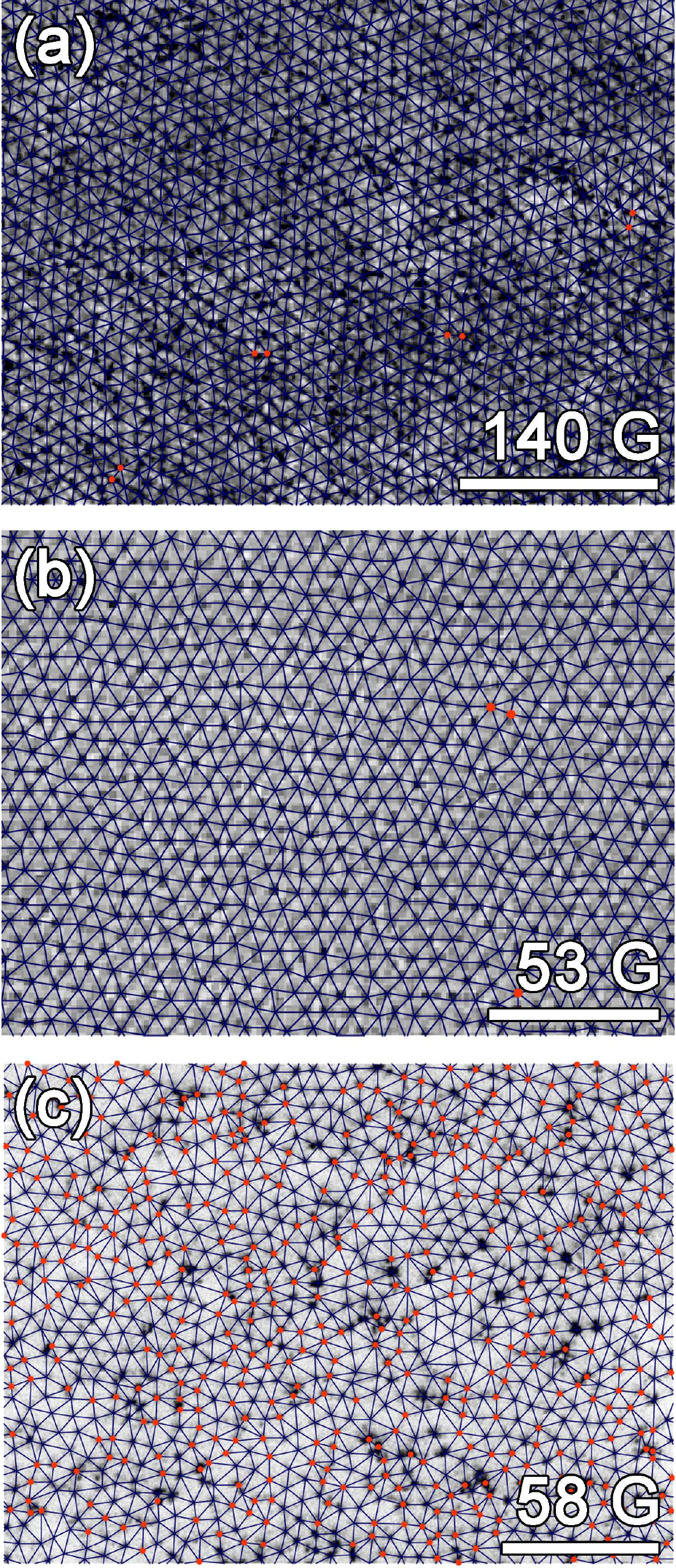}
    \caption{Magnetic decoration images of the vortex structure (black dots)
    nucleated in Bi$_2$Sr$_2$CaCu$_2$O$_{8+\delta}$
    samples with different types of disorder: (a) pristine sample
    with point pins; (b) electron-irradiated sample with extra point pins;
    (c) heavy-ion irradiated sample with a low density of columnar defects
    ($B_{\Phi}=30$\,Gauss, Xe-irradiated). The magnetic induction
    $B$ controlling vortex density is indicated in every panel. In
    all cases the white bar corresponds to 5\,$\mu$m. Each
    decoration image has overimposed Delaunay triangulations joining
    near-neighbor vortices with blue lines. Non-sixfold coordinated vortices are
    highlighted in red.}
    \label{fig:decos}
\end{figure}

\begin{figure}[ttt]
\includegraphics[width=0.65\columnwidth]{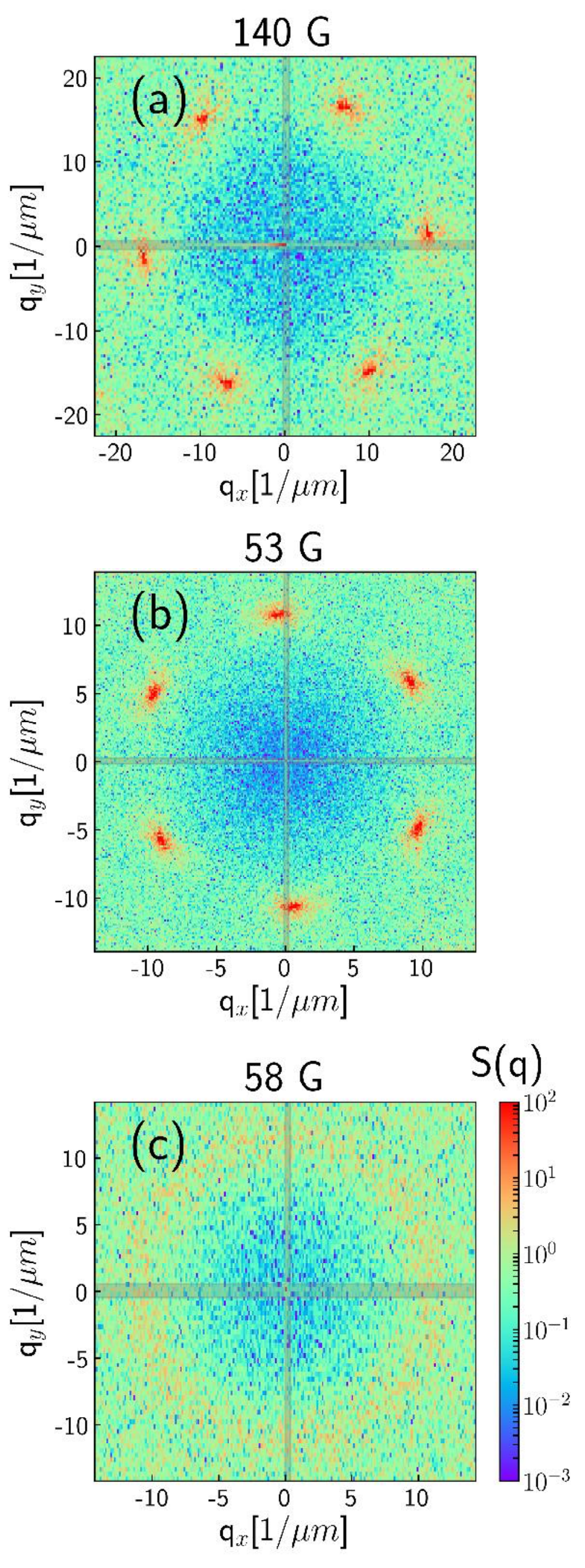}
\caption{Two-dimensional structure factors $S(q_x,q_y)$  for the
examples of vortex structures shown in Fig.\,\ref{fig:decos}. The
intensity is shown in a logarithmic scale and the color level is the
same for the three images, see color bar at the bottom.  The gray
crosses indicate the $q$-window affected by spurious effects arising
from the field-of-view edges. Data in these crosses are not
considered for the calculation of the angularly-averaged structure
factor $S(q)$.} \label{fig:Sofq}
\end{figure}

\begin{figure*}[ttt]
 \includegraphics[width=2\columnwidth]{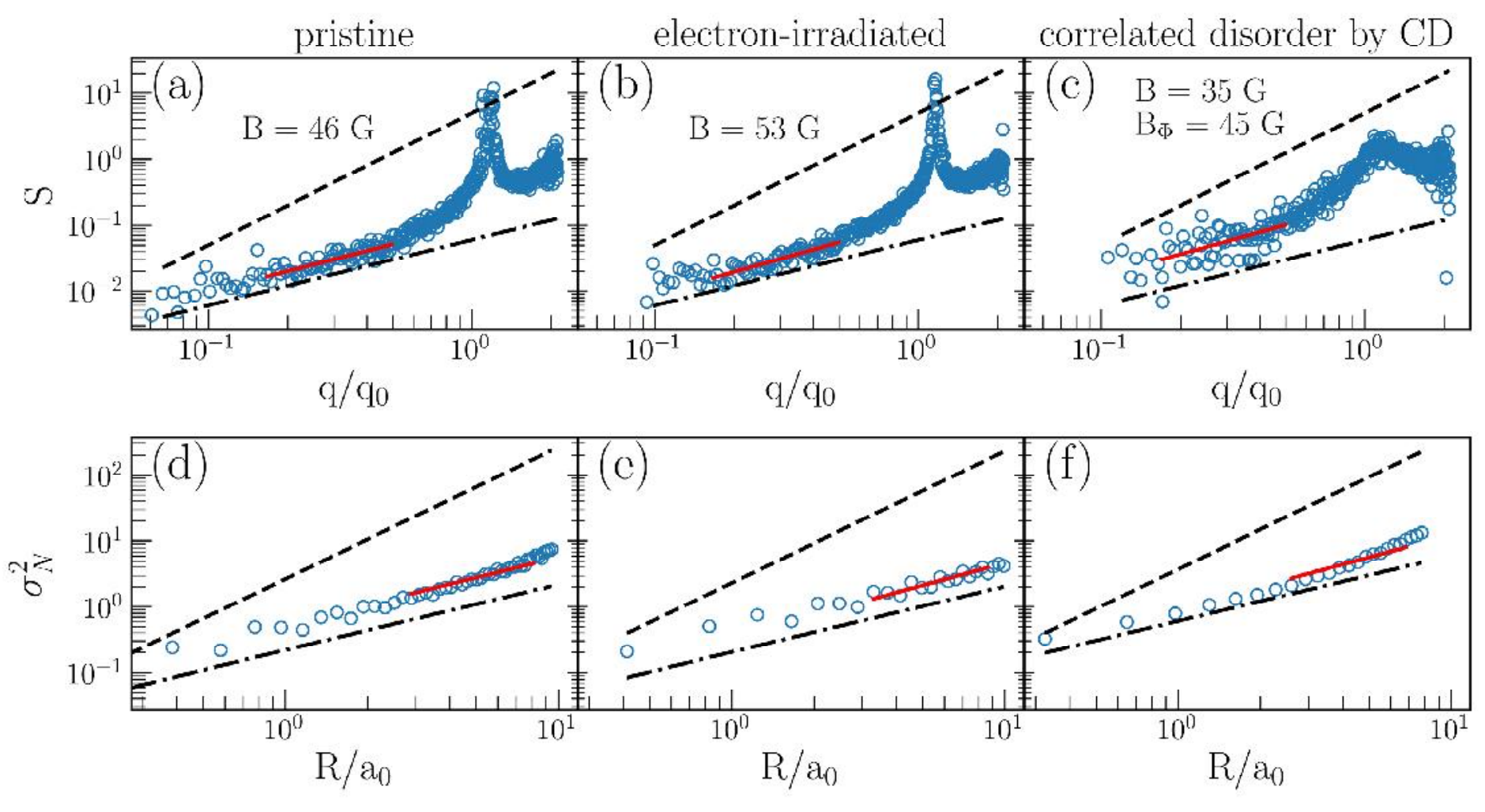}
 \caption{Angular average of the structure factor for the
 magnetically decorated vortex structure at the surface of (a) pristine,
 (b) electron-irradiated, and (c) Xe-irradiated (correlated CD
 disorder) Bi$_2$Sr$_2$CaCu$_2$O$_{8+\delta}$ samples. The vortex density in
 every case is indicated. Data are
 shown as a function of $q/q_0$ with $q_0 =2\pi/a_0$ the Bragg wave-vector.
 The red line is the best power-law fit $S(q) \sim q^\alpha_{\rm eff}$ for the low
  wave-vector range $a_{0}/6 <q/q_0 < a_{0}/2$.
 Number variance $\sigma_{N}^2$ as a function of
$R/a_0$ corresponding to the the same samples of the top panel: (d)
pristine, (e) electron-irradiated, and (f) Xe-irradiated samples.
The red line shows functions  $\sim R^{2-\alpha_{\rm eff}}$ with the
$\alpha_{\rm eff}$ obtained from fitting the $S(q)$ data of the top
panels. The dashed-line with slope $\alpha=2$ and the dashed-dotted
line with   slope $\alpha=1$ are guides to the eye.}
 \label{fig:sofqvsvariance}
\end{figure*}

We image individual vortex positions in a typical field-of-view of
thousands of vortices by performing magnetic decoration experiments
at 4.2\,K after field cooling at different applied fields in the
range $5<H<150$\,Oe. Vortices are decorated with Fe particles
attracted by the local field gradient generated around the vortex
cores~\cite{Fasano1999}, observed as black dots in the inverted
scanning-electron-microscopy images of Fig.\,\ref{fig:decos}.
Further details in the field-cooling decoration protocol followed in
this case can be found in Ref.\,~\onlinecite{Fasano1999}. For every
studied sample, several ($\sim 10$) magnetic decoration experiments
were performed in freshly cleaved surfaces, eventually at different
applied fields. We studied $\sim 30$ pristine, 1 electron-
irradiated, and 10 heavy-ion irradiated single crystals.

Figure~\ref{fig:decos} show examples of magnetic decoration images
at various applied fields in the three types of studied samples.
Panels (a) and (b) correspond to snapshots of vortex positions taken
at a field well within the quasi-crystalline Bragg glass phase for
pristine and electron- irradiated samples. For these two types of
samples, at fields larger than 15\,Gauss, the vortex structure is
single-crystalline, presents quasi-long-range positional order and
very few topological defects associated with non-sixfold coordinated
vortices. This is observed in the Delaunay triangulations
superimposed to the pictures with blue lines joining first
neighbors. Topological defects are highlighted with red dots. For
both images only $\sim 2$\,\% of vortices are involved in defects,
mainly edge dislocations. For fields smaller than 15\,Gauss, the
structure breaks into small crystallites for
pristine~\cite{Fasano1999} as well as electron- irradiated samples.
This polycrystalline structure results from vortex-vortex
interaction weakening and disorder becoming more relevant on the
viscous freezing dynamics~\cite{Fasano1999}. Further details on the
field-evolution of the structural properties (images, density of
defects, $S(q)$, displacement correlator and correlation lengths) of
the Bragg glass phase in pristine and electron- irradiated samples
can be found in our previous work Ref.\,\onlinecite{Jazmin2019}.

The structural properties of the vortex matter nucleated in samples
with CD are qualitatively different than for the case of point
disorder: small misaligned crystallites with less than 20 vortices
are observed at densities up to 100\,Gauss, see
Fig.\,\ref{fig:decos}, and for fields smaller than $\sim 15$\,Gauss
the structure is amorphous. In the latter case the density of
topological defects can be as high as 50\,\%, whereas at high fields
the amount of non-sixfold coordinated vortices is always larger than
40\,\%. The nucleation of these structures with at best 10 lattice
spacing short-range positional order comes from the strong effect of
pinning introduced by CD that are randomly distributed in the
sample~\cite{Menghini2003}. Nevertheless, the positional correlation
of these vortex structures is not that of a random distribution of
points mimicking the Poisson-like distribution of CD: the pair
correlation function of the vortex structure presents one very good
defined principal peak at $r=a_{0}$ and secondary weak peaks in some
cases.

\begin{figure*}[ttt]
 \includegraphics[width=2\columnwidth]{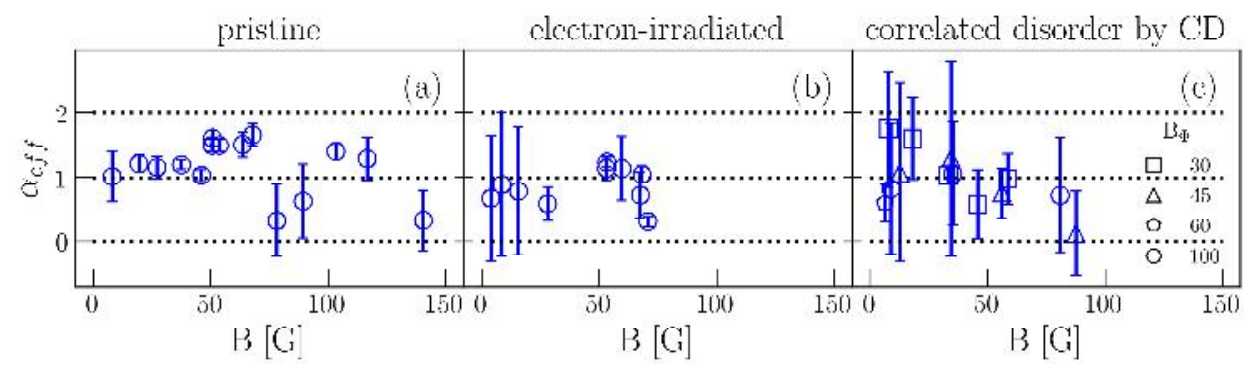}
 \caption{Effective exponents  obtained
 fitting  the structure factor  as $S(q) \sim q^{\alpha_{\rm eff}}$
 for the  magnetically decorated vortex structures in (a) pristine,
 (b) electron-irradiated and (c) heavy-ion irradiated samples
  as a function of vortex densities $B$.}
 \label{fig:allsamples}
\end{figure*}

The decorated structures were frozen, at length scales of order
$a_{0}$, at temperatures at which the pinning generated by disorder
sets in~\cite{Pardo1997,Fasano2005}. This freezing temperature is
some K below the temperature at which magnetic response becomes
irreversible~\cite{CejasBolecek2016}. This lenght-scale is shorter
than the relevant for the hyperuniformity analysis. Due to their
different relaxation rates, density modes corresponding to
length-scales larger than $a_0$ are expected to freeze at higher
temperatures than the modes with $q=2\pi/a_{0}$. We will further
discuss this issue later.

The images of Fig.\,\ref{fig:decos} are just some examples, but
typically we study a panoramic images of the vortex structure with
no less than 1500 and up to 15000 vortices. We digitalize vortex
positions in these large fields-of-view and then we calculate the
structure factor $S(q)$ and the number variance $\sigma_{N}^{2}$.
The larger the field-of-view, the smaller the value of $q\equiv
\sqrt{q_x^2+q_y^2}$ that we can access to calculate the
two-dimensional structure factor at the top surface. Figure
\ref{fig:Sofq} show the corresponding two-dimensional structure
factors $S(q_x,q_y)$ for the examples of vortex structures shown in
Fig.\,\ref{fig:decos}.    The $S(q_x,q_y)$ values are angularly
averaged as to obtain the $S(q)$ data shown in
fig.\,\ref{fig:sofqvsvariance}.

We would like to point out some important technical difficulties
that are quite specific to the study of the low-$q$ density modes.
On one hand, the borders and shape of our field-of-view hinders the
study of the structure factor in the relevant low-$q$ range due to
the annoying windowing effect. In rectangular fields-of-view the
artifact shows up as an excess in $S(q_x,q_y)$ localized in a
``$+$''-shaped region centered at $q_x=q_y=0$. This artifact,
associated to the Fourier transform of the field-of-view, is
oriented along the principal directions of the rectangle and has an
oscillatory decay on increasing $q$. This border effect is avoided
in simulations with in-plane periodic boundary conditions. To get
rid of this spurious effect, for the smallest $q$ we perform a
partial average over $\Psi$-values outside the `$+$''-shaped region,
instead of averaging over all $\Psi$. This is possible due to the
high angular localization of the artifact. Nevertheless, since the
cross has a finite width, at the end we are forced to consider a
safety minimal wave-vector $q_{\tt min}>2\pi/W_{fov}$, with
$W_{fov}$ a characteristic linear size of the field-of-view. On the
other hand, as we do not have an ensemble of many
magnetically-decorated vortex configurations at exactly the same
position of the samples to average over, inevitably, the lower the
$q$, the bigger the statistical fluctuations in $S(q)$. Henceforth,
in practice, although our images span distances as large as
$W_{fov}\sim 50 a_0$, we end up analyzing density fluctuations with
$\qperp > q_{\tt min} \gtrsim q_{0}/6= 1/6(2\pi/a_0)$.

Typical results for $S(q)$ are shown in
Fig.\,\ref{fig:sofqvsvariance} (top panels) for the pristine (a),
electron- irradiated (b), and heavy-ion irradiated  (c) samples. The
dashed and dashed-dotted lines are guides to the eye, showing $q^2$
and $q$ evolutions, respectively. For the lowest $q/q_0$ the
behavior is consistent with  class-II uniformity with $\alpha=1$ for
the three samples. Indeed, red lines in the figure shows a fit to
the data yielding $\alpha_{\rm eff} \sim 1$. Another way to analyze
the density fluctuations is to study the distance-evolution of the
vortex-number variance $\sigma_{N}^{2}$, (defined in
Eq.\,\ref{eq:variancedef}). To this end, we considered circular
regions with radius $R$ and centers located at random within the
panoramic image of the vortex structure. We pay attention to the
circles not crossing or touching the edge of the field-of-view. We
make statistics on the number of vortices contained over a large
amount of circles of size $R$ and the number variance
$\sigma_{N}^{2}(R)$ is then computed. Fig.\,\ref{fig:sofqvsvariance}
(bottom panels) shows the results of the $R$-evolution of
$\sigma_{N}^2$ corresponding exactly to the structures considered in
the top panels. We observe a rough agreement with $S(q)$, in the
sense that a class II hyperuniform scaling $\sigma_{N}^2 \sim R/a_0$
 fairly describes the data (see the dashed-dotted line). Indeed, the
 red lines in Figs.\,\ref{fig:sofqvsvariance} (d), (e) and (f)
 correspond to functions $\sim R^{2-\alpha_{\rm eff}}$ (see Eq.\,\ref{eq:variancedef})
  with the $\alpha_{\rm eff}$ obtained from the fits of the $S(q) \sim q^{\alpha_{\rm eff}}$
  data of the top panels.

 In order to perform a comprehensive study of the occurrence of
 hyperuniformity in vortex matter with different types of disorder and
 for different vortex densities,
 we have systematically fitted
$S(q)\sim q^{\alpha_{\rm eff}}$ for all a our studied samples and
vortex densities (magnetic field). The effective power-law exponent
$\alpha_{\rm eff}$ is obtained by fitting
 in the range $q/q_0 \in [1/2,1/6]$. The effective exponents as a
 function of field for every type of disorder are shown in the three
 panels of Fig.\,\ref{fig:allsamples}, compiling data from over roughly 40
statistically independent cases. Although $S(q)$ and $\sigma_{N}^2$
qualitatively agree, we have found that a systematic fit of
$\alpha_{\rm eff}$ using the expected scaling of $\sigma_{N}^2$  is
empirically more difficult due to the strong non-asymptotic
corrections to the number variance~\cite{Torquato2018}. As we can
observe,  $\alpha_{\rm eff} \approx 1 \pm 0.3$ is rather robust for
all the studied cases, independently of the type of disorder,
correlated or uncorrelated, present in the samples.

\section{\label{sec:discusion}Discussion and Perspectives}

A strict hyperuniformity analysis requires data in an asymptotic
regime which is rather difficult to reach and assure experimentally.
Nevertheless, our experimental results in an extensive data-set
display a clear suppression of the amplitude of the density
fluctuations with $S(q)\sim 10^{-2}$ for the lowest $q$ accessed (to
be compared with $S(q)=1$ for the Poisson or ideal gas particle
distribution). Hence, the vortex structures frozen during the
decoration field-cooling protocol are nearly hyperuniform
two-dimensional point patterns at the superconductor surface.
Furthermore, these structures display $S(q) \sim q^{\alpha_{\rm
eff}}$ with $\alpha_{\rm eff} \approx 1 \pm 0.3$ in the low-$q$
range accessed in our experiments. This agrees with the equilibrium
hydrodynamic predictions for the large-scale fluctuations of the
liquid phase with weak uncorrelated disorder, as well as for the
solid and liquid vortex phases in the ideal clean case. However,
this exponent contrasts with the predictions for the Bragg glass
phase ($\alpha=0$) and for the  liquid and Bose glass phases with CD
($S(q) \to \text{const}$ as $q \to 0$). We will argue  that these
discrepancies can be explained by considering the unavoidable
relevance of memory effects during the decoration field-cooling
protocol that affect the observed vortex structures at different
length-scales. These non equilibrium effects are ignored in the
hydrodynamic description of Sec.\,\ref{sec:theory}. We will also
argue that dispersion effects and finite-size effects are
experimentally relevant for a systematic study of density
fluctuations in vortex matter.

The two-dimensional point pattern obtained at the surface of the
sample after a field-cooling magnetic decoration down to 4.2\,K
corresponds to an out-of-equilibrium structure. This is due to the
strong dependence of the relaxation rate of the vortex structure
with the wavelength of the density modes.  In field-cooling
experiments with a fixed cooling rate, large length-scale density
fluctuations  have a slower relaxation rate than smalllength-scale
ones, and then the former fall out of equilibrium at larger
temperatures. This out-of-equilibrium effect is expected to be
enhanced by disorder that dramatically slows down the
thermally-activated dynamics. In other words, a magnetic decoration
image of the vortex structure at low  temperatures is not just a
snapshot but also a photograph with memory of its field cooling
history. Then, the freezing temperature of the vortex structure is
not a unique magnitude but rather depends on the length-scale of the
density fluctuations, $T_{\rm freez} = T_{\rm freez}(q)$. In a
field-cooling experiment, at $T \sim T_{\rm freez}(q)$ the density
fluctuations of mode $q$ have a relaxation rate of the order of the
experimental cooling rate. Namely, at $T <T_{\rm freez}(q)$ all
modes with wavelength $\leq 1/q$ are out of equilibrium.

In particular, the local ordering  at the scale $a_0$, observed in
samples with weak uncorrelated disorder  indicates that $T_{\rm
freez}(q \sim a_{0}^{-1}) \leq T_m$, with $T_m$ the melting
temperature of the Bragg glass phase. Density fluctuations
associated to larger wavelengths are then expected to fall out of
equilibrium at much higher temperatures, $T_{\rm freez}(q \ll
a_{0}^{-1}) > T_m$. These slow modes with $q \to 0$ retain memory of
the liquid phase. This can explain our experimental observation of a
near class-II hyperuniform vortex structure  in pristine and
electron-irradiated samples (predicted in the equilibrium
hydrodynamic approximation for the liquid phase with weak disorder).
This argument can be further justified in the framework of
disordered elastic systems without topological defects. Indeed, even
a simple elastic string relaxing in a random medium after a
temperature quench displays a logarithmically growing correlation
length~\cite{Blatter1994,Kolton2005} separating equilibrated from
non-equilibrated length-scales. Glassy vortex dynamics thus prevents
the system to reach the marginal hyperuniformity predicted for the
Bragg glass phase at equilibrium. This argument also explains why it
has been in general difficult to observe the different crossover
regimes predicted for the Bragg glass in magnetic decoration
experiments. Indeed, decoration experiments have been so far only
capable of revealing the random manifold, but not the
random-periodic regime of lattice roughness in samples with weak
uncorrelated disorder~\cite{Kim1999,CejasBolecek2016}.

The above out-of-equilibrium qualitative explanation is more subtle
to apply for the samples with CDs correlated disorder since in this
case the line-liquid as well as the Bose glass are expected to be
non-hyperuniform at equilibrium in an hydrodynamic approximation.
How is then possible to obtain $\alpha_{\tt eff}\approx 1$ from the
memory of a liquid with correlated disorder? A possible answer can
come from the crossover of density fluctuations at $q_{CD}$ from a
class-II hyperuniform liquid to a
 non-hyperuniform liquid on decreasing $q$. Observing the equilibrium
 structure at $q \to 0$ implies measuring
 at very large length scales $\sim 1/q_{CD}$ that might be hard to
 access experimentally in finite fields-of-view.
To evaluate this hypothesis we need to compute
\begin{equation}
\qperp_{CD} \approx \frac{n_0^2 \left(U_{0}^{2} b_{0}^{4} /
d^{2}\right)}{{\tilde c_{11}} k_B T_{\rm freez}(q_{CD})}
\sqrt{\frac{{\tilde c_{44}}}{{\tilde c_{11}}}}
\end{equation}
\noindent where we have used in Eq.\,\ref{eq:qcd} that $\Delta_{1}
\approx \left(U_{0}^{2} b_{0}^{4} /
d^{2}\right)\left[1+\mathcal{O}\left(b_{0}^{2} /
d^{2}\right)\right]$ is the disorder correlator strength for CD with
radius $b_0$ and mean separation $d \gg b_0$. It is difficult to
make a quantitative assessment of $q_{CD}$ due to the many
microscopic parameters involved. However, correlated disorder by CD
is expected to slow down the dynamics with respect to the weak
disorder case, and then to  increase $T_{\rm freez}(q)$ for all $q$.
Since $q_{CD} \propto 1/T_{\rm freez}(q_{CD})$, then $q_{CD}$ can
become too small to be experimentally accessed for our typical
fields-of-view. Thus, the observed structures effectively display a
class-II hyperuniformity due to the memory of the liquid phase, with
the first term of Eq.\,\ref{eq:Sofq2dboseliquid} dominating for
$q>q_{CD}$ at very high temperatures. A systematic study is
desirable to check this hypothesis, for instance comparing different
densities of CD. In particular, increasing $\Delta_{1}$ will produce
an enhancement of $q_{CD}$, and then the predicted saturation of
$S(q)$ due to the dominance of the correlated disorder term might be
observable in the same field-of-view.


\begin{figure}[h]
 \includegraphics[width=0.8\columnwidth]{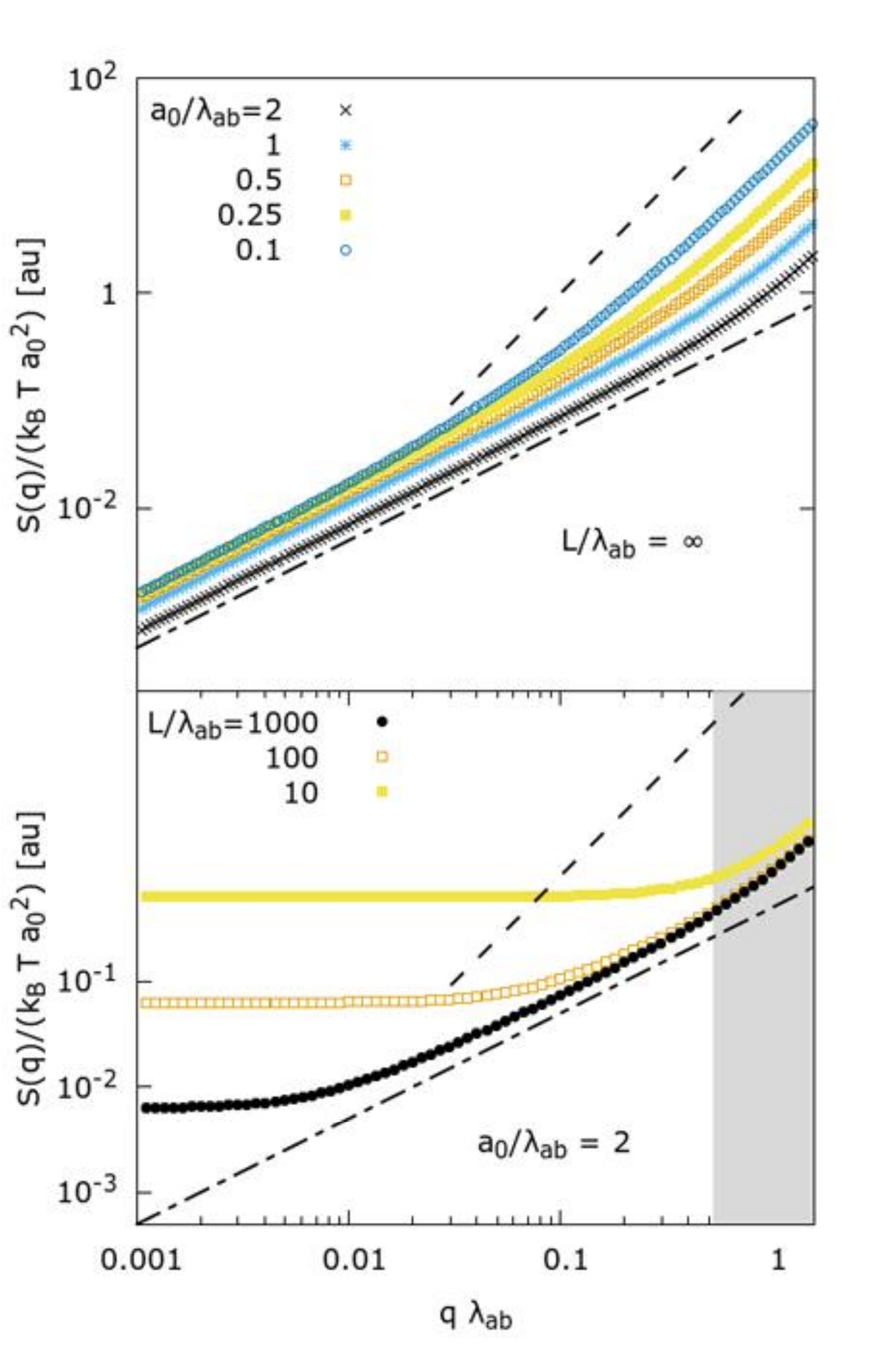}
 \caption{Constant-$z$ cross-section structure factor $S(q)$
 for the three dimensional line-liquid without disorder.
Top panel: Infinitely thick sample $L/\lambda_{ab} \gg 1$ for
various $a_0/\lambda_{ab}$ corresponding to different applied
magnetic fields. At high fields, corresponding to the smallest
$a_0\lambda_{ab}$ values, a crossover from the type II hyperuniform
scaling  $S(q)\sim q$ to $S(q)\sim q^2$ is observed at
 $q \lambda_c \approx 1$ (dotted lines are a guide to the eye for the two regimes).
 At low fields, the crossover
 shows up at larger $q$. The largest value of $a_0/\lambda_{ab}=2$ (black curve)
 roughly corresponds to the typical field of 40\,G applied in magnetic decorations
 for Bi$_2$Sr$_2$CaCu$_2$O$_{8 + \delta}$ with $\lambda_{ab}(T_{\rm freez})=0.4$\,$\mu$m.
 Bottom Panel: Finite-thickness effect in the structure factor
 $S(q)$ for various $L/\lambda_{ab}$ ratios at a fixed field such that $a_0/\lambda_{ab}=2$.
This effect produces the saturation of $S(q)$ in a region of small
$q$ whose extension depends on $L/\lambda_{ab}$.
   The case of
 $L/\lambda_{ab}=1000$ (black points) corresponds to a Bi$_2$Sr$_2$CaCu$_2$O$_{8 + \delta}$
  sample with a thickness of  400\,$\mu$m, a typical value in real
 samples. The gray shaded area  corresponds to the $q \cdot \lambda_{ab}$ range accessed
  experimentally  for our typical field-of-view in decorations (obtained considered
 again $\lambda_{ab}(T_{\rm freez})=0.4$\,$\mu$m). Finite size effects
 in $S(q)$ are not dominant for the experimental window accessed in
 our experiments in Bi$_2$Sr$_2$CaCu$_2$O$_{8 + \delta}$. However,
 dispersivity effects do play a role in our experimental window.}
 \label{fig:dispersion}
\end{figure}

Another effect that we have so far not considered in the discussion
is the anisotropy-enhanced dispersivity of the elastic constants for
the lowest wave-vectors accessed in our fields-of-view. This is
particularly important in layered materials as
Bi$_2$Sr$_2$CaCu$_2$O$_{8 + \delta}$ since dispersivity can induce,
as discussed, a crossover from class-II hyperuniformity to a
different behavior on increasing $q$. Since we access scales $\qperp
\cdot \lambda_{ab} \ll 1$, in-plane dispersivity is not important.
However, dispersion in the $z$-direction is important when $q \cdot
\lambda_{c} = \Gamma q \cdot \lambda_{ab} \sim 1 $.  The anisotropy
parameter of our samples is $\Gamma \approx 170$, and then the
condition $q \cdot \lambda_{c} \sim 1$ can easily  be reached in our
experimental fields-of-view since $\lambda_c (T_{\rm freez}) =
\Gamma \lambda_{ab}(T_{\rm freez}) \approx 70$\,$\mu$m. For an
anisotropic superconductor the compression modulus is
\cite{Brandt1995}

\begin{equation}
c_{11}(q,q_{z})=\frac{B^{2}}{4\pi} \frac{1+\lambda_{c}^{2} (q^{2} +
q_{z}^{2})}{\left(1+\lambda_{a b}^{2} (q^{2}+
q_{z}^{2})\right)\left(1+\lambda_{c}^{2} \qperp^{2}+\lambda_{a
b}^{2} q_{z}^{2}\right)} \label{eq:c11bisco}
\end{equation}
\noindent while the tilt modulus reads
\begin{equation}
c_{44}(q,q_{z})=\frac{B^{2}}{4\pi}\left[\frac{1}{1+\lambda_{c}^{2}
\qperp^{2}+\lambda_{a b}^{2} q_{z}^{2}}\right] + c'_{44}(q_{z})
\label{eq:c44bisco}
\end{equation}
\noindent with $c'_{44}(q_z)$ the isolated-vortex contribution. This
last term, important at low fields, is dispersive in $q_z$  for
small wave-length fluctuations only, such that $q_z \lambda_{ab}
\sim 1$. This dispersivity is due to the electromagnetic coupling
between pancake vortices~\cite{Blatter1994}. For the long
wave-length fluctuations we are interested in, we can  take the
constant non-dispersive limit for the isolated vortex contribution,
\begin{equation}
c'_{44}
\approx
\frac{B^2}{4\pi} \left(\frac{a_0}{\lambda_{ab}}\right)^2
\end{equation}
\noindent If we assume $L \to \infty$ and integrate
Eq.\,\ref{eq:Sofq3dliquid} over $q_z$ using the expressions of
Eqs.\,\ref{eq:c11bisco} and \ref{eq:c44bisco}, we obtain the
two-dimensional structure factor $S(q)$ including dispersion
effects. The result for our Bi$_2$Sr$_2$CaCu$_2$O$_{8 + \delta}$
samples is shown in Fig.\,\ref{fig:dispersion}, for different vortex
densities in the case of an infinite sample, see top panel.

The bottom panel of Fig.\,\ref{fig:dispersion} shows $S(q)$ for
different sample thicknesses at a fix vortex density
$a_{0}/\lambda_{ab}(T_{\rm freez})=2$ corresponding to a magnetic
induction of $\sim 40$\,G.  We considered $\lambda_{ab}(T_{\rm
freez}) \sim 0.4$\,$\mu$m at the irreversibility temperature
$T_{irr}(B \sim 40\,\text{Gauss})$. This is the temperature at which
pinning sets in, and the vortex structure is frozen at length-scales
of $a_{0}$~\cite{CejasBolecek2016}. The gray area indicates the $q
\cdot \lambda_{ab}(T_{\rm freez})$ range used for fitting
$\alpha_{\rm eff}$ in our experimental $S(q)$ data.

In the case of an infinite sample,
 $S(\qperp)$ presents a crossover, from $S(\qperp) \sim
\qperp$ to $S(\qperp) \sim q^2$ at larger $q \cdot \lambda_{ab}$.
This large-$q$ deviation from the asymptotic hyperuniform class-II
behavior for clean samples is due to the dispersivity in the elastic
constants.  The $q$-location of the crossover decreases on
increasing field. In particular, in the limit of high fields, where
the isolated vortex contribution $c'_{44}$ can be neglected, the
crossover scale tends to $\lambda_c (T_{\rm freez})$, see top panel
of Fig.\,\ref{fig:dispersion}. Then, in order to ascertain whether a
structure is hyperuniform, it is crucial to obtain $\alpha_{\tt
eff}$ by fitting $S(q)$ data in a $q$-range located below this
dispersivity-induced crossover.

In addition, this fit can neither be performed in the low-$q$ limit
where finite size effects due to the finite sample thickness $L$
destroy hyperuniformity. The bottom panel of
Fig.\,\ref{fig:dispersion}  shows that this finite size effect,
observed as a saturation of $S(q)$ at low $q$ values, extends over a
larger $q$-range on decreasing $L$. This was already discussed in
Sec.\,\ref{sec:finitesizeeffects}, though neglecting the
dispersivity and strong anisotropy of  the elastic constants for
vortex matter in Bi$_2$Sr$_2$CaCu$_2$O$_{8 + \delta}$. The figure
highlights a black curve corresponding to the prediction for our
approximate experimental situation, namely vortex density $\sim
40$\,Gauss and $L \sim 400$\,$\mu$m. In our experiments, the
saturation associated to the finite size effect is well below the
$q$-range taken into account in the $\alpha_{\rm eff}$ fits
(gray-shaded area). Therefore our hyperuniformity study in vortex
matter from magnetic decoration images is not affected by this
finite-size effect. However, for our experimental length-scales, the
saturation should be clearly visible for samples with $L /
\lambda_{ab}(T_{\rm freez}) \approx 10$, namely for $L \approx
4$\,$\mu$m. This is a rather thin sample to be obtained just by
cleaving a thicker crystal. Magnetic decoration data in 1\,$\mu$m
thick Bi$_2$Sr$_2$CaCu$_2$O$_{8 + \delta}$ micron-sized samples are
available, but the field-of-view is of only a few hundred of
vortices, making difficult to perform a careful hyperuniformity
analysis of such vortex
nanocrystals~\cite{Dolz2015,CejasBolecek2016}.

Going back to the relevance of dispersivity effects for our
hyperuniformity analysis, the departure from the  $S(q) \sim q$
behavior for our typical experimental situation, see black curve in
Fig.\,\ref{fig:dispersion}, starts at the upper half of our
experimental fitting window. Therefore, although the easily
cleavable Bi$_2$Sr$_2$CaCu$_2$O$_{8 + \delta}$ system makes the
experimental realization of this magnetic decoration study possible,
the hyperuniformity analysis is messed up by the dispersivity
effects inherent to its extremely-layered nature. Then the
$\alpha_{\rm eff}$ values obtained in our study might be
slightly-overestimated due to this effect.

In order to study up to what point our obtained $\alpha_{\rm eff}$
values are overestimated due to out-of-equilibrium and/or
dispersivity effects, further experiments should be performed.
First, out-of-equilibrium effects could be quantified by altering
the field-cooling process, either by significantly slowing down the
cooling rate, or by adding an in-plane dithering field allowing the
vortex system to more efficiently relax towards the free energy
minimum~\cite{Zeldovdithering}. Second, dispersivity effects can be
reduced by choosing a less anisotropic host superconducting
material. In a more general perspective, complementary experimental
studies of vortex density fluctuations by applying other different
techniques, such as neutron scattering, would be of great interest.
From the theoretical point of view, extending the hydrodynamic
theory to the realistic non-equilibrium situation of field-cooling
experiments, as well as performing non-equilibrium simulations under
the same conditions, would be very insightful.

In summary, we have shown that a directed three-dimensional elastic
line array with short-ranged interactions in all directions can
generate, at equilibrium, a two-dimensional hyperuniform point
pattern at their cross sections. This behavior can be explained by
the three-dimensional bulk energetics involved in the compression of
a single layer, and ultimately in the continuity of lines. We have
shown  that disorder can play an important role on modifying in
several interesting ways such  hyperuniform behavior. In order to
test the above physical picture we have chosen the case-study system
of vortex structures in type-II superconductors, a paradigmatic
experimental realization of the three-dimensional elastic line
array. We have studied this issue both, theoretically via analytic
calculations and numerical simulations, and experimentally in the
anisotropic vortex matter nucleated in Bi$_2$Sr$_2$CaCu$_2$O$_{8 +
\delta}$  samples with correlated and uncorrelated disorder.  By
magnetically decorating the frozen two-dimensional vortex point
pattern at the top surface of different superconducting samples we
find a systematic suppression of vortex-density fluctuations at long
wavelengths. We have argued that this is roughly consistent with the
phenomenological hydrodynamic predictions if we assume that the
density modes of the observed configurations keep memory of the
liquid-state at the largest length-scales probed by the structure
factor. We have also discussed the effect of non-asymptotic
behaviors, such as the ones arising from the dispersivity of elastic
constants and non-equilibrium effects, and finite-size effects.

The kind of study we have done may offer a different viewpoint to
analyze vortex phases, their transitions, and also to characterize
the somewhat unavoidable non-equilibrium relaxation effects in
disordered systems. This includes not only static vortex phases as
the ones we have analyzed in this work, but also dynamical vortex
phases such as current-driven
lattices~\cite{Pardo1998,Ledoussal1998,Balents1998,Scheidl1998,NattermannScheidl2000,LeDoussal2010},
where genuine non-equilibrium stationary effects may add interesting
extra ingredients to the density fluctuations. In all cases, it
would be interesting to analyze whether new hyperuniform patterns
emerge, either at the level of the whole system or in a subsystem
(as the constant-$z$ cross section we study here), and to check if
they share some of the interesting response properties discussed in
Ref.~\onlinecite{Torquato2018} for general hyperuniform states of
matter.

We acknowledge illuminating discussions with C. Olson-Reichhardt, P.
Le Doussal, T. Giamarchi and H. Suderow.

\bibliography{biblio}

\end{document}